\documentclass[%
reprint,
superscriptaddress,
nofootinbib,
amsmath,amssymb,
aps,
prc,
prc,
floatfix, ]%
{revtex4-2}

\usepackage{graphicx}
\usepackage{dcolumn}
\usepackage{bm}
\usepackage[bookmarks=true,colorlinks,%
            citecolor=blue,linkcolor=blue,anchorcolor=blue,filecolor=blue,urlcolor=blue,%
           ]{hyperref}
\allowdisplaybreaks

\newcommand{\bra}[1]{\langle {#1} |}
\newcommand{\ket}[1]{| {#1} \rangle}
\newcommand{\inproduct}[2]{\langle #1 | #2 \rangle}

\begin{document}

\title{
Fermi operator expansion method for nuclei and inhomogeneous matter
	with nuclear energy density functional
}

\author{Takashi Nakatsukasa}%
 \affiliation{Center for Computational Sciences,
              University of Tsukuba, Tsukuba 305-8577, Japan}
 \affiliation{Faculty of Pure and Applied Sciences,
              University of Tsukuba, Tsukuba 305-8571, Japan}
 \affiliation{RIKEN Nishina Center, Wako 351-0198, Japan}

\date{\today}

\begin{abstract}
\noindent{\bf Background:}
The nuclear energy density functional method at finite temperature
is a useful tool for studies of nuclear structure at high excitation,
and also for researches of nuclear matter involved in
explosive stellar phenomena and neutron stars.
However, its unrestricted calculation
requires large computational costs for the three-dimensional coordinate-space solvers,
especially for the Hamiltonian matrix diagonalization
and (or) the Gram-Schmidt orthonormalization of the single-particle wave functions.
\\
\noindent{\bf Purpose:}
We test numerical performance of a numerical method,
that requires neither the diagonalization nor the Gram-Schmidt orthonormalization,
for finite nuclei and inhomogeneous nuclear matter.
We examine its advantageous features in future applications.
\\
\noindent{\bf Methods:}
The Fermi operator expansion method,
which approximates the Fermi-Dirac distribution in terms of
the Chebyshev polynomials,
is used to construct the one-body density matrix
for the energy density functional calculations at finite temperature.
The modified Broyden's mixing method is adopted for
the self-consistent iteration process.
\\
\noindent{\bf Results:}
The method is applied to isolated finite $N=Z$ nuclei and to non-uniform symmetric
nuclear matter at finite temperature,
which turns out be very effective with the three-dimensional coordinate-space
representation, especially at high temperature.
The liquid-gas transition is clearly observed in the calculations.
\\
\noindent{\bf Conclusions:}
The Fermi operator expansion method is a useful tool for
studies of various nuclear phases at finite temperature
with the energy density functional calculations.
The method is suitable for massively parallel computing with distributed memory.
Furthermore, when the space size is large,
the calculation may benefit from its order-$N$ scaling property.
\end{abstract}


\maketitle

\section{Introduction}

It is of significant importance to calculate nuclear matter
in a variety of phases with different temperature,
utilized in simulation studies of supernovae and neutron stars.
The nuclear energy density functional method at finite temperature
\cite{BR86,Sch19}
is a desirable choice for studying
the inhomogeneous neutron-star matter in outer and inner crusts.
Especially, near the boundary between the inner crust and the core,
various exotic phases, ``nuclear pasta'', are expected to appear.

In order to properly treat thermally dripped nucleons and to study
the transition from inhomogeneous to uniform nuclear matter,
the coordinate-space representation
is preferable.
Furthermore, to find a new exotic structure at finite temperature,
it is desired to perform the calculation without assuming
any spatial symmetry of the configuration,
using the three-dimensional (3D) coordinate-space representation.
Since the 3D coordinate-space solution
is computationally demanding,
most of the finite-temperature mean-field calculations for nuclei
either adopt
the harmonic-oscillator-basis (shell-model-basis) representation
\cite{Goo81,ER93,BM16,ZN18},
or are restricted to the spherical systems\footnote{
An exception can be found in a paper by Newton and Stone \cite{NS09}
in which they solve the finite-temperature Hartree-Fock equations
with the BCS treatment on the pairing correlation
in the 3D coordinate space.
To reduce the computational time,
the states with the occupation smaller than $10^{-6}$ are neglected.
}
\cite{BLV84,BLV85}.

To reduce the computational cost,
the finite-temperature Thomas-Fermi approximation has been often adopted
\cite{BGH85,OPP97,OMYT13,XMYT22}.
The molecular dynamics simulations can be performed
with even smaller computational time, thus,
they have been extensively utilized with larger simulation volumes
\cite{WSYE04,Hor15,CH17}.
A major drawback of these semiclassical approximation is a lack of shell effects.
In the inner crust, the shell effects play a role not only
in protons but also in the band effect for unbound neutrons
scattered by the periodic potentials \cite{BM01,Cha05}.
An alternative quantum approach to the inner crust is
to use the Wigner-Seitz approximation,
pioneered by Negele and Vautherin \cite{NV73}.
The structure is optimized in a Wigner-Seitz sphere of radius $R_\mathrm{\sc ws}$.
using different boundary conditions depending on the parity
of the single-particle orbitals.
For the inner crust,
this trick for the boundary condition produces a roughly constant
neutron density at the spherical boundary $r=R_\mathrm{\sc ws}$.
However, some spurious density fluctuations still remain near the edge
of the boundary.
Furthermore, the numerical results suffer from ambiguity
caused by the choice of the boundary conditions \cite{BST06}.
It should be noted that
a combination of the Thomas-Fermi and Wigner-Seitz approximations
is frequently used for non-uniform matter at finite temperature
based on the microscopic results for the uniform matter
\cite{STOS98,Tog17,XMYT22}.
Another disadvantage of these approximations is
that one loses information on the transport properties
which may be crucial for understanding
dynamics of neutrons in the inner crust of neutron stars
\cite{CH08}.

In this paper, we perform a feasibility study for
the fully quantum energy density functional (mean-field) calculations
without the Wigner-Seitz approximation
for non-uniform nuclear matter at finite temperature.
A conventional solution of the finite-temperature mean-field theory
can be summarized as follows:
(1) Construct the mean-field Hamiltonian $H$ which depends on
one-body densities.
(2) Diagonalize the Hamiltonian to obtain the eigenvalues
and the eigenvectors, $H\ket{i}=\epsilon_i\ket{i}$.
(3) Calculate the densities, then, go back to (1)
to reach the self-consistency.
In the step (3), the Fermi-Dirac distribution function $f(x)$ is
used to calculate the densities, $\rho=\sum_i f(\epsilon_i)\ket{i}\bra{i}$.
The truncation with respect to the eigenvector $\ket{i}$
may be possible at low temperature,
while, at high temperature,
we need to compute all the eigenvalues and eigenvectors.
Since this diagonalization is needed every iteration,
it requires a large amount of numerical resources.

Recently, the shifted Krylov method for the 
Hartree-Fock-Bogoliubov (HFB) theory has been proposed \cite{JBRW17}.
Then, it was extended to the finite-temperature HFB theory \cite{KN20}.
The method uses the shifted Krylov subspace method for
solution of linear algebraic equation, $(z-H)G(z)=0$,
where $H$ is the HFB Hamiltonian and $G(z)$ is the Green's function.
The densities are obtained from the Green's function $G(z)$ integrated
over complex energy $z$.
Thus, the diagonalization procedure is unnecessary.
This feature is favorable for large systems since the matrix
diagonalization needs the operation of O$(N^3)$ where
$N$ is the dimension of the matrix.
It is shown to be numerically feasible and efficient
in the parallel computation \cite{JBRW17,KN20}.
However, its performance depends on the required number of
iterations of the shifted Krylov algorithm
whose convergence is not guaranteed.

Purposes of the present paper are to study
an alternative method for the finite-temperature mean-field calculation,
and to examine its performance for nuclear systems.
The methodology is known as the Fermi operator expansion (FOE) method
in the condensed matter physics \cite{GC94,GT95}.
It is also known as one of the order-$N$ (O($N$)) method \cite{WJ02},
thus, the number of computational operations linearly scales with respect to
either the particle number or the dimension of the one-particle space.
In the O($N$) methods,
the ``nearsightedness'' of many electron systems play a crucial role \cite{Kohn95}.
Since the nearsightedness is due to destructive interference effect
in quantum mechanical many-particle systems,
we expect that it is applicable to nuclear systems as well.
However, since the size of a nucleus is roughly ten femtometer at most,
the nearsightedness principle has been assumed not so beneficial in practice.
The situation may be different for hot nuclei and macroscopic neutron-star matters.
It is worth examining the O($N$) methods
for calculations of nuclei at finite temperature and
inhomogeneous nuclear matter.

The paper is organized as follows:
The finite-temperature mean-field theory is recapitulated in Sec.~\ref{sec:FTMFT}.
In Secs.~\ref{sec:FOE} and ~\ref{sec:Chebyshev},
the Fermi operator expansion method is summarized.
In Sec.~\ref{sec:entropy},
we propose an efficient method of computing the entropy
without calculating single-particle energies.
The nearsightedness and the O($N$) method are briefly reviewed
in Sec.~\ref{sec:order-N}.
Details of the numerical calculations,
examination of the validity of the Chebyshev polynomial expansion,
and numerical results for finite nuclei and non-uniform matter
are shown in Sec.~\ref{sec:results}.
Concluding remarks are given in Sec.~\ref{sec:conclusion}.

\section{\label{sec:theory} Theory and numerical methods}


\subsection{\label{sec:FTMFT} Mean-field theory at finite temperature}

We recapitulate here the Hartree-Fock (HF) theory at finite temperature
\cite{BR86}.
The partition function and the statistical density matrix
at the temperature $\beta^{-1}=k_B T$ are 
in the form, $Z=\mathrm{tr}\left[ e^{-\beta \hat{H}'} \right]$,
and $\hat{D}=e^{-\beta \hat{H}'}/Z$, respectively,
where $\hat{H}'\equiv \hat{H}-\mu \hat{N}$ with the one-body HF Hamiltonian
$\hat{H}$, the particle number operator $\hat{N}$, and the chemical potential $\mu$.
The one-body density matrix is given as
\begin{equation}
	\rho_{ij}=\mathrm{tr}\left[ \hat{D} \hat{c}_j^\dagger \hat{c}_i \right]
	= \sum_\alpha \inproduct{i}{\alpha}f_{\beta\mu}(\epsilon_\alpha)\inproduct{\alpha}{j}
	\label{eq:rho_ij}
\end{equation}
where $f_{\beta\mu}$ is the Fermi-Dirac function
$f_{\beta\mu}(x)=\{1+e^{\beta (x-\mu)}\}^{-1}$.
Here, the subscripts $i$ and $j$ denote the indices for an arbitrary
single-particle basis,
while $\alpha$ for the single-particle states to diagonalize $H$ (and $H'$),
$\hat{H}\ket{\alpha}=\epsilon_\alpha\ket{\alpha}$.
$\hat{c}_i$ ($\hat{c}_i^\dagger$) is an annihilation (creation)
operator for a particle at the state $\ket{i}$.
Since the HF Hamiltonian $\hat{H}[\rho]$ is a functional of
the one-body density, the states $\ket{\alpha}$ and energies $\epsilon_\alpha$
depend on $\rho_{ij}$.
Thus, Eq.~(\ref{eq:rho_ij}) should be iteratively calculated
until the self-consistency is achieved.
It is straightforward to extend the theory to
the HFB theory at finite temperature
\cite{BR86,KN20}.

It should be worth mentioning that
the finite-temperature HF theory
can be derived by the principle of maximum entropy,
with an assumption that the partition function is given in a form,
$Z=\mathrm{tr} [ e^{-\beta \hat{K}} ]$ with
a one-body operator $\hat{K}$.
Constraining the energy and the particle number
with Lagrange multipliers (associated with $\beta$ and $\mu$),
it is equivalent to
the minimization of the thermodynamic potential
\cite{BR86}.
\begin{eqnarray}
	J&=& E - TS - \mu N \\
	&=& E[\rho]
	+ k_B T \mathrm{tr} \left[ \hat{D}\ln\hat{D} \right]
	-\mu \mathrm{tr} \left[ \hat{D}\hat{N} \right] \\
	&=& E'[\rho] - \mathrm{tr} \left[ \hat{D}\hat{K} \right]
	- k_B T \ln Z ,
\end{eqnarray}
where $E[\rho]$ is the energy density functional and
$E'\equiv E-\mu N$.
Taking the variation with respect to the one-body operator $\delta\hat{K}$,
it leads to
\begin{eqnarray}
	\delta J &=& \frac{\delta E'}{\delta\rho}\cdot\delta\rho
	-  \mathrm{tr} \left[ \hat{D}\delta\hat{K} \right]
	-  \mathrm{tr} \left[ \hat{K}\delta\hat{D} \right] 
	-\frac{\delta \ln Z}{\beta} \\
	&=& \mathrm{tr} \left[ \hat{H}' \delta\hat{D} \right]
	- \mathrm{tr} \left[ \hat{K}\delta\hat{D} \right] ,
\end{eqnarray}
where $\hat{H}'\equiv\sum_{ij} H'_{ij} \hat{c}_i^\dagger \hat{c}_j$
with $H'_{ij}=\delta E'/\delta\rho_{ji}$.
Therefore, the principle of maximum entropy gives $\hat{K}=\hat{H}'$.

\subsection{\label{sec:FOE} Fermi operator expansion method}

According to Eq. (\ref{eq:rho_ij}),
the one-body density can be calculated
by diagonalizing $H'$
to obtain the eigenstates and the eigenenergies, $\ket{\alpha}$ and $\epsilon_\alpha$.
However, since we need to perform the diagonalization
every iteration until the self-consistency is achieved,
it is prohibitively difficult for large systems.
In order to reduce the computational cost,
we must avoid the matrix diagonalization
which numerically costs O$(N^3)$.
In this paper, we explore one of such approaches,
the Fermi operator expansion (FOE) method.

The idea of the FOE method can be easily understood by
rewriting Eq. (\ref{eq:rho_ij}) as
$\rho_{ij}=\bra{i}\hat{\rho}\ket{j}$,
where
\begin{equation}
	\hat{\rho}=\sum_\alpha \ket{\alpha} f_{\beta\mu}(\epsilon_\alpha)\bra{\alpha}
	= f_{\beta\mu}(\hat{H}) .
	\label{eq:rho}
\end{equation}
Thus, the one-body density is nothing but the Fermi-Dirac function whose
argument is replaced by the Hamiltonian.
In addition, the FOE is based on the polynomial approximation
of the Fermi-Dirac distribution function.
\begin{equation}
	f_{\beta\mu}(x) \approx \sum_{n=0}^M a_n T_n(x) , 
	\label{eq:polynomial_approximation}
\end{equation}
where $T_n(x)$ is a polynomial function of the $n$-th degree,
and the summation is truncated at the maximum degree $M$.
The polynomial approximation should be better at large $T$,
because the Fermi-Dirac function is smoother at higher temperature.
In contrast, at the zero temperature limit,
the function becomes the Heaviside step function for which
the approximation is not so precise.
Nevertheless, in case that there is a gap $\Delta E$ at the Fermi surface,
such as the shell gap and the pairing gap,
the results of the finite temperature calculation with $k_B T\ll\Delta E$
is practically identical to the one at zero temperature.

Inserting Eqs. (\ref{eq:rho}) and (\ref{eq:polynomial_approximation})
into $\rho_{ij}=\bra{i}\hat{\rho}\ket{j}$,
we have
\begin{equation}
	\rho_{ij}=\bra{i} f_{\beta\mu}(\hat{H}) \ket{j}
	=\sum_{n=0}^M a_n \inproduct{i}{j_n},
	\label{eq:rho_ij_approx}
\end{equation}
where $\ket{j_n}\equiv T_n(\hat{H}) \ket{j}$.
If the polynomial function $T_n(x)$ is simply given by $T_n(x)=x^n$,
the state $\ket{j_n}$ can be calculated starting from $\ket{j_0}\equiv\ket{j}$
as
\begin{equation}
	\ket{j_n} =\hat{H} \ket{j_{n-1}},
	\quad n=0,\cdots,M.
	\label{eq:recursive_relation}
\end{equation}
Thus, multiplying the basis state $\ket{j}$ by $\hat{H}$ $M$ times,
the one-body density $\rho_{ij}$ can be constructed.
This is the basic idea of the FOE method.

In practice, the simple choice of $T_n(x)=x^n$ often leads to
a numerical instability, because the functions $x^n$ are diverging function
at $|x|>1$ for large $n$.
In order to avoid this numerical problem,
a careful choice of the polynomial functions is required for $T_n(x)$.

\subsection{\label{sec:Chebyshev}Chebyshev polynomials}

In the present work, we adopt the Chebyshev polynomials for $T_n(x)$.
The Chebyshev polynomials of the first kind are given by
$T_n(x)=\cos nt$ with $x=\cos t$,
thus, both $x$ and $T_n(x)$ are bound between $-1$ and $1$.
They are orthogonal with respect to the weight of $1/\sqrt{1-x^2}$.
\begin{equation}
	\int_{-1}^1 T_n(x) T_m(x) \frac{dx}{\sqrt{1-x^2}} = N_n \delta_{nm} ,
\end{equation}
with the normalization constants $N_0=\pi$ and $N_n=\pi/2$ ($n\neq 0$).

First, we should change the energy scale by
transforming $\hat{H}$ into
$\mathbf{\hat{H}}\equiv \frac{\hat{H}-e_c}{e_r}$
where $e_c\equiv (e_{\rm max}+e_{\rm min})/2$ and
$e_r\equiv (e_{\rm max}-e_{\rm min})/2$.
When the eigenvalues of $\hat{H}$ satisfy
$e_{\rm min}\leq e_\alpha \leq e_{\rm max}$
in the adopted model space,
those of $\mathbf{\hat{H}}$ are in the interval $[-1,1]$.
Instead of expanding $f_{\beta\mu}(x)$ as Eq. (\ref{eq:polynomial_approximation}),
we expand a scaled Fermi-Dirac function $\tilde{f}(x)$ as
\begin{equation}
	\tilde{f}(x) \equiv f_{\beta\mu}(e_r x + e_c)
	\approx \frac{a_0}{2}+\sum_{n=1}^M a_n T_n(x) ,
	\label{eq:polynomial_approximation_2}
\end{equation}
where the coefficients $a_n$ are given by
\begin{equation}
a_n=\frac{2}{\pi}\int_{-1}^1 T_n(x) \tilde{f}(x) \frac{dx}{\sqrt{1-x^2}} .
	\label{eq:a_n}
\end{equation}
It is worth noting that $\tilde{f}(x)$ and $a_n$
depend on both $\beta$ and $\mu$,
for which we omit these subscripts for simplicity.

Instead of Eq. (\ref{eq:recursive_relation}),
the recursive relations of the Chebyshev polynomials,
\begin{equation}
	T_{n+1}(x)=2x T_n(x) - T_{n-1}(x) ,\quad n\geq 1,
	\label{eq:recursive_relation_Chebyshev}
\end{equation}
lead to recursive formula for $\ket{j_n}\equiv T_n(\hat{\mathbf{H}})\ket{j}$,
\begin{equation}
	\ket{j_{n+1}} =2 \mathbf{\hat{H}} \ket{j_n} - \ket{j_{n-1}},
	\quad n=1,\cdots,M-1.
	\label{eq:recursive_relation_2}
\end{equation}
Starting with $\ket{j_0}=\ket{j}$ and
$\ket{j_1}=\mathbf{\hat{H}} \ket{j}$,
all the states $\ket{j_n}$ up to $n=M$ are obtained,
then, the one-body density is calculated as
\begin{equation}
	\rho_{ij}
	=\bra{i} \tilde{f}(\hat{\mathbf{H}}) \ket{j}
	=\frac{a_0}{2}\inproduct{i}{j}+\sum_{n=1}^M a_n \inproduct{i}{j_n}.
	\label{eq:rho_ij_approx_2}
\end{equation}

Let us summarize the numerical procedure to reach the self-consistent solution
of the HF problem at a given temperature $T$.
\begin{enumerate}
\setcounter{enumi}{-1}
	\item\label{step:setup}
		The maximum and minimum energies, $e_{\rm max}$ and $e_{\rm min}$,
		are determined according to a problem of interest.
		See Sec.~\ref{sec:numerical_details} for those values.
		The initial density distribution $\rho(\mathbf{r})$ and
		the initial chemical potential $\mu$ are given by hand.
		For a given $T$, calculate the coefficients $a_n$
		($n=0,\cdots,M$) according to Eq. (\ref{eq:a_n}).
		Set up the initial Hamiltonian $\hat{H}$.
	\item\label{step:cal_j}
		Calculate $\ket{j_n}$ ($n=0,\cdots,M$) according to
		Eq. (\ref{eq:recursive_relation_2}).
	\item\label{step:density}
		Construct the one-body density according to
		Eq. (\ref{eq:rho_ij_approx_2}).
		Adjust the chemical potential $\mu$ if necessary.
	\item\label{step:Hamiltonian}
		Construct the HF Hamiltonian $\hat{H}[\rho]$ using
		the calculated density $\rho_{ij}$.
	\item\label{step:check}
		Check the self-consistency between the density and the Hamiltonian.
		If it is self-consistent, end the iteration.
		Otherwise, go to Step \ref{step:cal_j} and iterate the procedure.
\end{enumerate}

In the present formulation,
the function $f_{\beta\mu}(x)$ depends on $T$ and $\mu$.
Thus, when we change the chemical potential $\mu$, 
we have to recalculate 
the coefficients $a_n$ in Eq. (\ref{eq:polynomial_approximation_2}).
You may think that it is better to expand the function
$(1+e^{\beta x})^{-1}$ instead of $\{1+e^{\beta(x-\mu)}\}^{-1}$ 
and to use $\hat{\mathbf{H}}-\mu$ instead of $\hat{\mathbf{H}}$.
However, in this case, we need reevaluate the states $\ket{j_n}$
using the recursion relation (\ref{eq:recursive_relation_2}),
because $\ket{j_n}$ depend on $T$ and $\mu$.
Since the calculation of $\ket{j_n}$ requires the major portion of the computation,
we adopt the expansion of $f_{\beta\mu}(x)$.
If we adjust the chemical potential in step \ref{step:density} of every iteration
to fix the particle number (average density),
only the coefficients $a_n$ in Eq. (\ref{eq:a_n}) need to be recalculated.
The additional computation is negligibly small
compared to the calculation of $\ket{j_n}$.

\subsection{\label{sec:entropy}Calculation of entropy}

It is of significant importance to calculate the entropy of systems
at finite temperature.
The calculation of the free energy requires the evaluation of
the entropy as well.
For the product wave functions, the entropy $S$ is given by
\begin{equation}
	S=-k_B \sum_\alpha \left[
		f_\alpha \ln f_\alpha
		+\left( 1-f_\alpha\right)
		\ln \left(1-f_\alpha \right) \right] ,
	\label{eq:S}
\end{equation}
where $f_\alpha=f_{\beta\mu}(\epsilon_\alpha)$.
In order to calculate this, normally we need all the eigenvalues
of the Hamiltonian , $\epsilon_\alpha$,
which requires an additional computation,
namely the diagonalization of the Hamiltonian.
It demands a large computational cost of O$(N^3)$.

In this paper, we propose another manner to approximate a function
\begin{equation}
	\tilde{g}(x)\equiv -\tilde{f}(x) \ln \tilde{f}(x)
	- \left\{1-\tilde{f}(x)\right\} \ln \left\{1-\tilde{f}(x)\right\} ,
\end{equation}
with the polynomial expansion as
\begin{equation}
	\tilde{g}(x) \approx \frac{b_0}{2}+\sum_{n=1}^{M'} b_n T_n(x) ,
\end{equation}
analogous to Eq. (\ref{eq:polynomial_approximation_2}).
The coefficients $b_n$ are determined in the same manner as
Eq. (\ref{eq:a_n}).
Then, the entropy can be calculated as
\begin{eqnarray}
	S&=&k_B {\rm tr}\left[\tilde{g}(\hat{\mathbf{H}}) \right] \\
	&\approx& \sum_j 
	\left[ \frac{b_0}{2}+\sum_{n=1}^{M'} b_n \inproduct{j}{j_n} \right] .
\end{eqnarray}
Since the states $\ket{j_n}=T_n(\hat{\mathbf{H}})\ket{j}$
are calculated in Eq. (\ref{eq:recursive_relation_2}) in order to construct the density,
almost no extra cost is needed for evaluation of the entropy provided that $M'\leq M$.
In fact, we find that the condition $M'\leq M$ is well satisfied
in practice (See Sec.~\ref{sec:validity}).
At small temperature,
$\tilde{g}(x)$ has a sharp peak at $x=(\mu-e_c)/e_r$,
which demands large value of $M'$.
However, in this case, $M$ must be also large, because
$\tilde{f}(x)$ also produces a sharp transition from 1 to 0.
At $T=0$, $\tilde{f}(x)$ becomes a discontinuous Heaviside function,
$\tilde{f}(x)=\theta(x)$, 
while $\tilde{g}(x)$ is a constant function, $\tilde{g}(x)=0$.

In the FOE method, we end up the vectors, $\ket{j_n}$,
which contains the information of the Chebyshev polynomials of the Hamiltonian,
$\ket{j_n}=T_n(\hat{\mathbf{H}})\ket{j}$.
Therefore, quantities that are continuous functions of
the single-particle energies,
including the density and the entropy,
can be evaluated in principle from $\ket{j_n}$.
The number of the vectors $\ket{j_n}$ is $N\times M$,
where $N$ is the dimension of the single-particle space (system size).
Since $M$ is inversely proportional to the temperature $T$ as Eq. (\ref{eq:M})
below,
the FOE is more efficient at higher temperature.

\subsection{Nearsightedness and order-$N$ method}
\label{sec:order-N}

The FOE method is regarded as one of the linear system-size scaling method,
namely, the order-$N$ (O($N$)) method.
According to Ref.~\cite{BH97},
the degrees of polynomials necessary for an accuracy of $10^{-D}$ ($D>1$)
is estimated as
\begin{equation}
	M= \frac{2}{3} (D-1) e_r \beta .
	\label{eq:M}
\end{equation}
Assuming a gaussian basis functions of range $\sigma$
centered at mesh points whose spacing comparable to $\sigma$,
the matrix elements for $\hat{H}$ at a large separation have the same width $\sigma$,
and the long-range matrix elements for $\hat{H}^M$ are estimated as
$\sqrt{M}\sigma$ \cite{BH97}.
Therefore, the range of the density matrix of Eq. (\ref{eq:rho_ij_approx_2})
is approximately given as
\begin{equation}
r_N\sim \sqrt{M}\sigma\sim \sqrt{\frac{\hbar^2}{3m}(D-1)\beta},
	\label{eq:r_N}
\end{equation}
where we use Eq. (\ref{eq:M}) and
$e_r \sim e_{\rm max} \sim \hbar^2\sigma^{-2}/(2m)$
at a small value of $\sigma$.
In other words, the density matrices $\rho(\mathbf{r},\mathbf{r}')$
are localized, namely, $\rho(\mathbf{r},\mathbf{r}')\approx 0$
at $|\mathbf{r}-\mathbf{r}'|>r_N$.
It becomes more ``nearsighted'' ($r_N\rightarrow 0$)
for higher temperature $\beta\rightarrow 0$.

At the zero temperature limit,
$r_N$ can stay finite if there is a gap $\delta e$ at the Fermi surface,
although Eq. (\ref{eq:r_N}) diverges.
Taking the chemical potential $\mu$ as the mid value of the gap,
the condition that discrepancies between the Heaviside function and
the Fermi-Dirac function are smaller than $10^{-D}$ except for
the gap interval leads to
\begin{equation}
	\beta > \left( 2 \ln 10\right) \frac{D}{\delta e} .
\end{equation}
For instance, when we have a shell gap of $\delta e=2$ MeV at the Fermi surface
and require the accuracy of $D=3$ (error smaller than $10^{-3}$),
the calculation at $T=100$ keV is practically identical to that at $T=0$.

The nearsightedness of the density enables us to perform the O($N$) calculation.
The calculation of $\ket{j_n}$ ($n=1,\cdots,M$)
in Eq. (\ref{eq:recursive_relation_2}) can be performed
in a truncated space whose dimension does not depend on the system size.
Since the nonlocal (off-diagonal) densities $\rho_{ij}$ with $R_{ij}>r_N$ vanish,
the matrix-vector product in Eq. (\ref{eq:recursive_relation_2})
can be performed in the restricted active subspace.
Here, $R_{ij}$ mean that spatial distance between two basis states
$\ket{i}$ and $\ket{j}$.
For the coordinate-space basis, they are trivially $R_{ij}=|\mathbf{r}_i-\mathbf{r}_j|$.
For the single-center harmonic-oscillator basis,
which are common and efficient in calculation of finite nuclei,
it is difficult to find a pair of states with $R_{ij}>r_N$.
Thus, the applicability of the O($N$) method also relies on the choice of the basis.

Before finishing this section, we emphasize
the following advantageous features of the method in numerical computation.
First of all, in order to construct the one-body density $\rho_{ij}$,
only the matrix-vector product,
operation of the Hamiltonian on a state,
is necessary in Eq. (\ref{eq:recursive_relation_2}).
Although the self-consistency between the density $\rho$ and
the Hamiltonian $H[\rho]$ requires the iteration,
neither the matrix diagonalization nor
the linear algebraic equations are involved to achieve the
mean-field solution at the temperature $T$.
Second, the calculations of $\rho_{ij}$ in Eqs. (\ref{eq:recursive_relation_2})
and (\ref{eq:rho_ij_approx_2}) for different $j$ can be
independently performed.
It is suitable for massively parallel computing for large systems.
Last, but not the least, 
the method may receive benefits from its nearsightedness,
and the computational cost could linearly scale with the system size
(Sec~\ref{sec:nearsightedness}).

\section{Numerical results}
\label{sec:results}

\subsection{Energy density functional and numerical details}
\label{sec:numerical_details}

We use the BKN energy density functional \cite{BKN76},
which is a functional of the isoscalar kinetic and local densities
and assumes the spin-isospin symmetry
without the spin-orbit interaction.
The pairing correlation is neglected.
Since the BKN functional is not suitable for description of the neutron-rich matter,
we study only the symmetric nuclear matter and finite nuclei with $N=Z$.
Nevertheless, it serves for the purposes
of the present paper,
namely, to examine applicability and performance
of the FOE method for the mean-field (energy-density) calculation for nuclei
and nuclear matter at finite temperature.

We adopt the 3D Cartesian grid representation \cite{NY05}
of the square box with periodic boundary condition.
The 3D grid size is set to be $h^3=(1.0\mbox{ fm})^3$.
The differentiation is evaluated with the nine-point finite difference.
For calculation of isolated finite nuclei,
the center-of-mass correction is taken into account by
modifying the nucleon's mass as $m\rightarrow m*A/(A-1)$.
For the non-uniform nuclear matter calculation ($A\rightarrow\infty$),
we use the bare nucleon's mass.
The fast Fourier transform is utilized for
calculation of the Coulomb potential,
which is well suitable for periodic systems.
For the calculation of the isolated finite nucleus,
we use the method same as Ref.~\cite{Sky3D}
following the idea given in Ref.~\cite{EB79}.

In order to make the Chebyshev polynomial expansion, we need to set
the maximum and the minimum single-particle energies.
The minimum energy is taken as $e_{\rm min}=-50$ MeV,
which is safe enough
in the cases of $N=Z$ nuclei and the symmetric nuclear matter.
The maximum energy is set as
\begin{equation}
e_{\rm max}=3\times \frac{\hbar^2}{2m}
	\left(\frac{\pi}{h}\right)^2,
\end{equation}
according to the maximum kinetic energy for the 3D grid of $h^3$.

For the self-consistent iteration of the finite-temperature
HF calculation,
we use the modified Broyden's method \cite{Joh88,Bar08}.
We use the HF potential $v(\mathbf{r})$
as the Broyden's vector to update \cite{Bar08}.

\subsection{Validity check for the polynomial expansion}
\label{sec:validity}

\begin{figure}[t]
	\centerline{
	\includegraphics[width=0.5\textwidth]{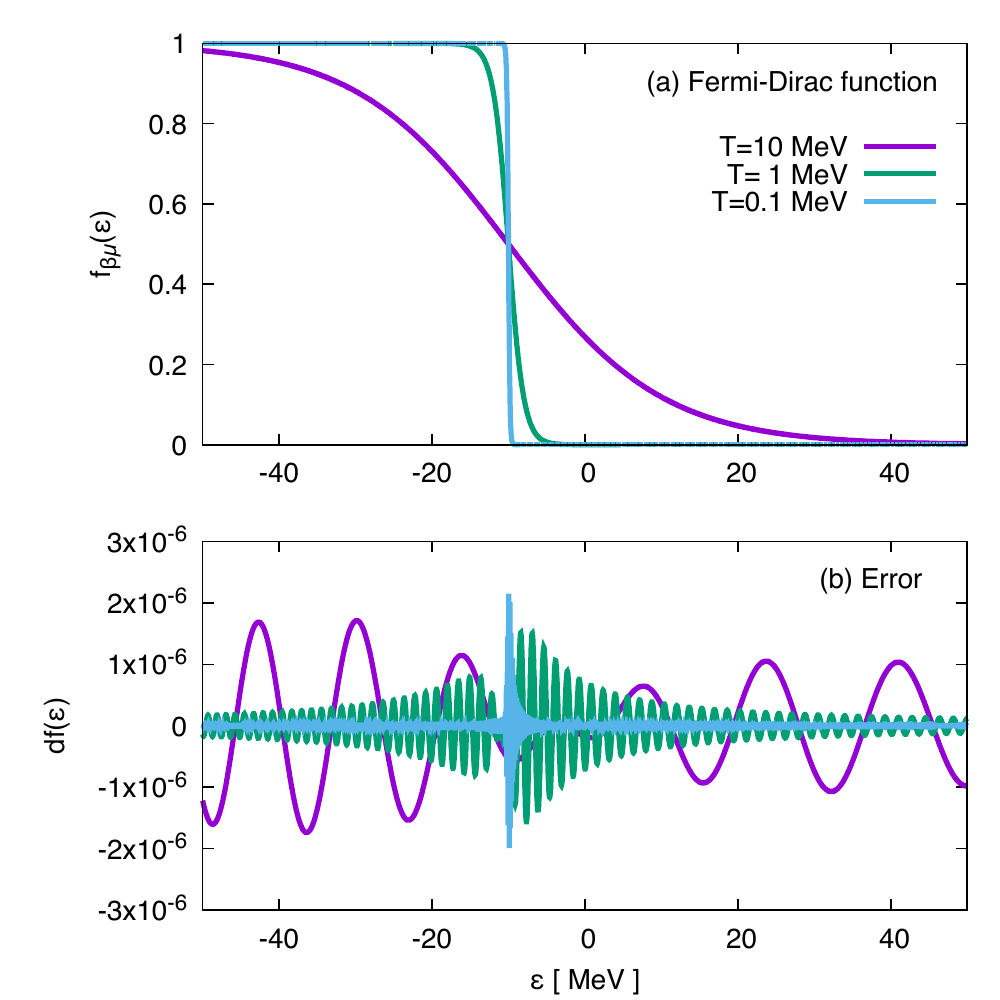}
}
	\caption{
		(a) Fermi-Dirac distribution function $f_{\beta\mu}(\epsilon)$
		as a function of the single-particle energy,
		calculated with the Chebyshev polynomial expansion.
		The degrees of polynomials are
		$M=90$, 900, 9,000 for $T=10$, 1, and 0.1 MeV, respectively.
		(b) Error in the polynomial approximation, Eq. (\ref{eq:df}).
	}
	\label{fig:FD}
\end{figure}

Let us first examine the accuracy of the expansion with the Chebyshev polynomials.
According to Eq. (\ref{eq:M}), the maximum degrees of the polynomials $M$
are adopted as $M=1.5 \times e_r \beta$,
which corresponds to the accuracy of $10^{-5.5}$.
In Fig.~\ref{fig:FD},
we show the approximated Fermi-Dirac distribution function $f_{\beta\mu}(\epsilon)$
with the chemical potential $\mu=-10$ MeV (panel (a)),
and the deviation from the exact values (panel (b)),
\begin{equation}
	df(\epsilon)=
	\frac{a_0}{2}+\sum_{n=1}^M a_n T_n\left(\frac{\epsilon-e_c}{e_r}\right)
	- f_{\beta\mu}(\epsilon) .
	\label{eq:df}
\end{equation}
We assume that the maximum and minimum single-particle energies are
$\epsilon_{\rm max}=500$ MeV and 
$\epsilon_{\rm min}=-100$ MeV, respectively,
which leads to $M=900 \beta$ ($\beta$ in units of MeV$^{-1}$).
The largest deviation appears around $\epsilon=\mu$ and
its value is order of $10^{-6}$, which is consistent with
the estimation of Eq. (\ref{eq:M}).
We can clearly see the importance of
the temperature-dependent maximum degrees $M$.

\begin{figure}[t]
	\centerline{
	\includegraphics[width=0.5\textwidth]{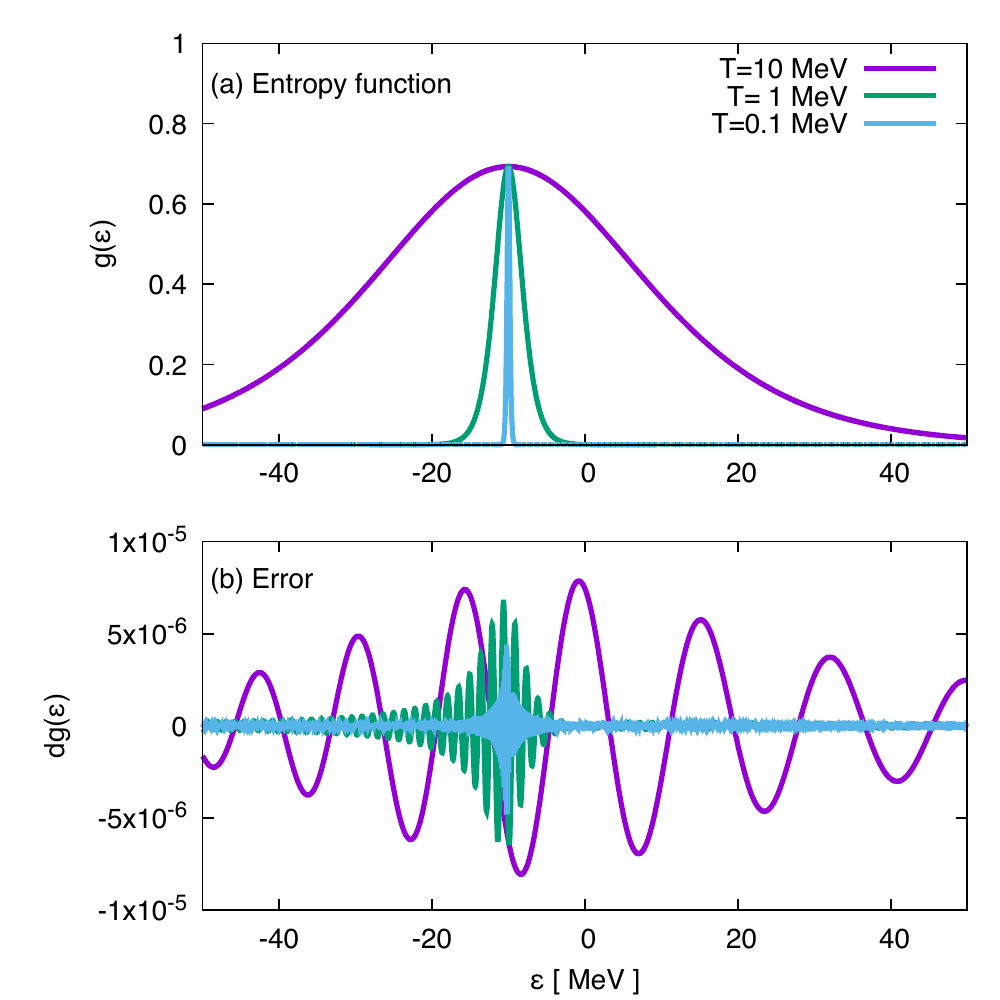}
}
	\caption{
		Same as Fig.~\ref{fig:FD}, but for
		the entropy function $g(\epsilon)$ of Eq. (\ref{eq:g})
		instead of $f_{\beta\mu}(\epsilon)$.
	}
	\label{fig:S}
\end{figure}

We perform the same analysis on the function, 
\begin{equation}
	g(\epsilon)= -f_{\beta\mu}(\epsilon) \ln f_{\beta\mu}(\epsilon)
	- \left\{1-f_{\beta\mu}(\epsilon)\right\} \ln \left\{1-f_{\beta\mu}(\epsilon)\right\} ,
	\label{eq:g}
\end{equation}
which is used for calculation of the entropy, Eq. (\ref{eq:S}),
and show the result in Fig.~\ref{fig:S}.
The maximum degrees $M'$ is taken as $M'=M=900\beta$.
The Chebyshev expansion for $g(\epsilon)$ is well approximated with
the deviation is smaller than $10^{-5}$ for any temperature.
They are well controlled as far as the maximum degrees $M$ are
adjusted in proportion to $\beta$.

\begin{figure}[t]
	\centerline{
	\includegraphics[width=0.5\textwidth]{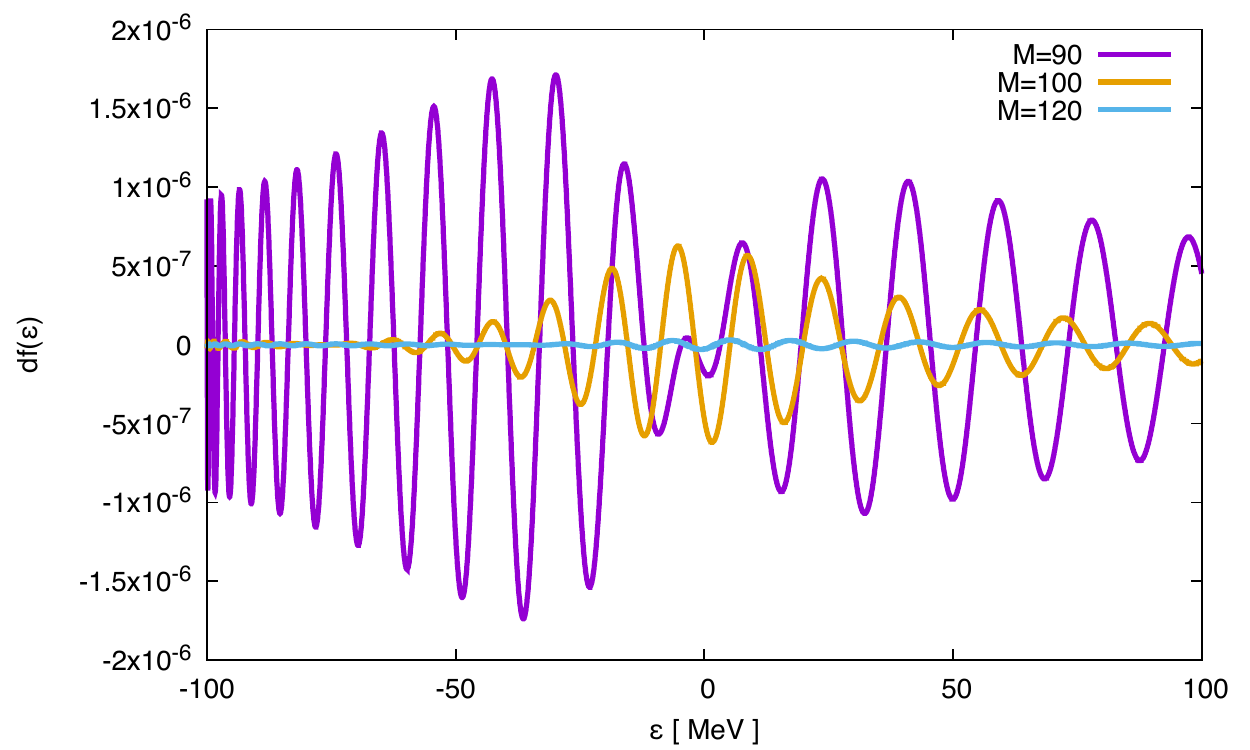}
}
	\caption{
		Error in the Fermi-Dirac function at $T=10$ MeV
		with different values of the maximum degrees $M$.
		The magenta line is calculated with $M=90$,
		the same as that in Fig.~\ref{fig:FD}(b).
		Those with $M=100$ and $M=120$ are shown by
		orange and light blue lines, respectively.
	}
	\label{fig:M_dep}
\end{figure}

If we increase $M$ at the fixed temperature,
the accuracy is significantly improved as shown in Fig.~\ref{fig:M_dep}.
We find an agreement with Eq. (\ref{eq:M});
$M=100$ ($M=120$) corresponds to $D=6$ ($D=7$) in Eq. (\ref{eq:M}).

In order to check the accuracy in the final results,
we perform the FOE calculation using
$M= k \times e_r \beta$ with
different values of $k$.
We use the Woods-Saxon potential,
$V_{\rm ws}(r)=V_{\rm ws}/(1+\exp(r-R_{\rm ws})/a_{\rm ws})$,
with $V_{\rm ws}=-50$ MeV, $R_{\rm ws}=3$ fm, and $a_{\rm ws}=0.5$ fm.
The adopted space size is $(13\mbox{ fm})^3$ and
the chemical potential is fixed at $\mu=-15$ MeV.
The calculated quantities with $T=1$ MeV and 10 MeV
are shown in Table~\ref{tab:accuracy}.
The nucleon number is calculated as the integration of density
over the adopted space.
The ``Woods-Saxon energy'' is defined as
$E_{\rm ws}=(1/2) \int V_{\rm ws}(\mathbf{r}) \rho(\mathbf{r}) d\mathbf{r}$.
The difference between $k=1.5$ and $k=2.0$
is negligible, less than 100 eV in energy.
In this paper, we use the temperature-dependent maximum degrees,
$M=1.5 \times e_r \beta$,
which provides a reasonable accuracy.

\begin{figure}[t]
	\centerline{
	\includegraphics[width=0.5\textwidth]{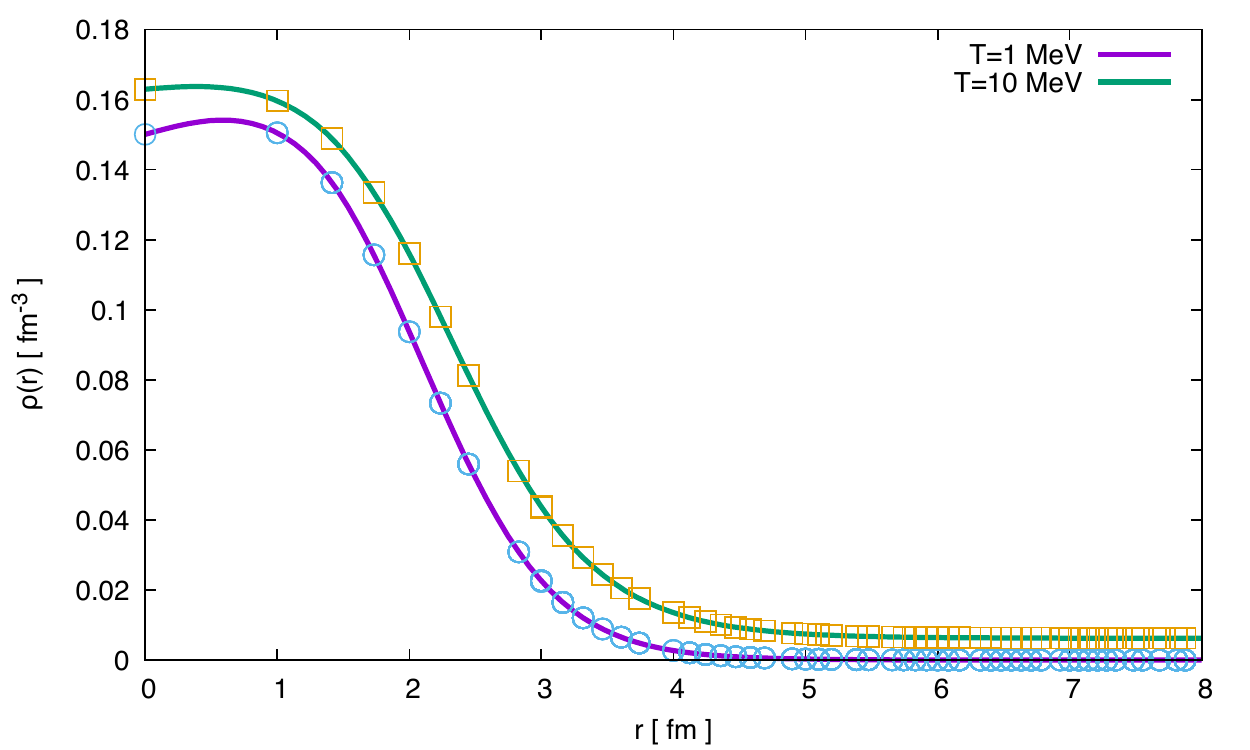}
}
	\caption{
		Calculated density profiles at $T=1$ MeV and 10 MeV
		as a function of the radial coordinate.
		Symbols indicate calculated values at the square mesh points
		and lines are obtained by the spline interpolation.
		See text for details.
	}
	\label{fig:rho_WS}
\end{figure}

With the same chemical potential,
the nucleon number $A$ increases approximately threefold
from $T=1$ MeV to 10 MeV.
On the other hand,
difference in the Woods-Saxon energy is only about 30\%.
This can be understood from the nucleon density profile
shown in Fig.~\ref{fig:rho_WS}.
A significant portion of nucleons is dripped from
the Woods-Saxon potential at $T=10$ MeV.
The particles far out of the potential range $R_{\rm ws}=3$ fm
do not contribute to $E_{\rm ws}$.

\begin{table}[b]
	\caption{
		Calculated values of nucleon number $A$, 
		the total kinetic energy $E_{\rm kin}$,
		the Woods-Saxon energy $E_{\rm ws}$,
		and the entropy $S/k_B$,
		using different values of $k$.
		The temperature is $T=1$ MeV for the upper three rows,
		while $T=10$ MeV for the rest.
		See text for details.
	}
\begin{ruledtabular}
\begin{tabular}{c|r r r r}
	$k$ & $A$ & $E_{\rm kin}$ [ MeV ] & $E_{\rm ws}$ [ MeV ] & $S/k_B$ \\
	\hline 
	& \multicolumn{4}{c}{$T=1$ MeV} \\
	1.0 & 9.8125 & 159.5155 & $-183.8843$ & 8.3121 \\
	1.5 & 9.8125 & 159.5157 & $-183.8843$ & 8.3121 \\
	2.0 & 9.8125 & 159.5157 & $-183.8843$ & 8.3121 \\
	\hline 
	\hline 
	& \multicolumn{4}{c}{$T=10$ MeV} \\
	1.0 & 27.3152 & 532.9096 & $-248.7950$ & 78.6638 \\
	1.5 & 27.3152 & 532.9055 & $-248.7954$ & 78.6634 \\
	2.0 & 27.3152 & 532.9055 & $-248.7954$ & 78.6634 \\
	\hline 
\end{tabular}
\end{ruledtabular}
	\label{tab:accuracy}
\end{table}

\subsection{Isolated nuclei at finite temperature}
\label{sec:isolated_nuclei}

\subsubsection{Modified Broyden's method}

In the HF calculations,
the self-consistency between the densities and
the potentials (Hamiltonian) is required.
For a given HF Hamiltonian $\hat{H}_{\rm in}^{(n)}$
at the $n$-th iteration,
the one-body density $\rho_{ij}$ is obtained
using the FOE calculation of Eq. (\ref{eq:rho_ij_approx_2}),
which produces a new HF Hamiltonian $\hat{H}_{\rm out}^{(n)}$.
The self-consistency is achieved when we reach the fixed point
for the Hamiltonian,
$H_{\rm in}=H_{\rm out}$,
which is equivalent to the density fixed point.
Since naive iteration with the total replacement of the Hamiltonian as
$\hat{H}_{\rm in}^{(n+1)}=\hat{H}_{\rm out}^{(n)}$
does not converges in many cases,
the linear mixing is often adopted as
$\hat{H}_{\rm in}^{(n+1)}=(1-\alpha)\hat{H}_{\rm in}^{(n)}
+\alpha\hat{H}_{\rm out}^{(n)}$
with a mixing parameter $\alpha$.
Although the divergence can be avoided
if we choose the parameter $\alpha$ small ($0<\alpha\ll 1$),
the convergence can be very slow.

\begin{figure}[t]
	\centerline{
	\includegraphics[width=0.5\textwidth]{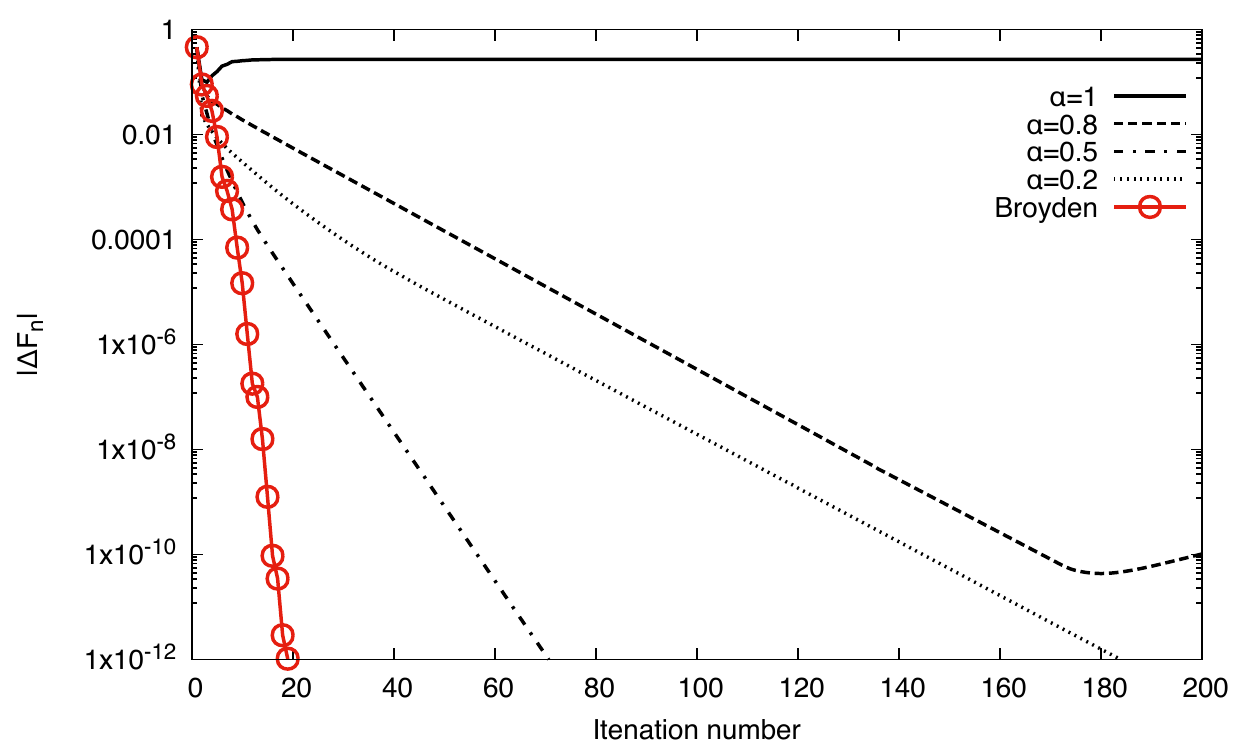}
}
	\caption{
		Comparison of convergence between the linear mixing (black lines)
		and the modified Broyden's method (red symbols) for $^{16}$O
		calculated at $T=1$ MeV.
		The space size is $(13\mbox{ fm})^3$.
		See text for details.
	}
	\label{fig:broyden}
\end{figure}

In this paper, we use the modified Broyden's method
\cite{Joh88}.
In Ref.~\cite{Bar08}, its performance has been examined for
finite nuclei with the Skyrme-HFB calculations at zero temperature
using the matrix diagonalization.
We perform similar study for the FOE calculation at
finite temperature.
In Fig.~\ref{fig:broyden},
we show the convergence behaviors of the modified Broyden's method,
compared to the linear mixing method.
Here, we show the difference in diagonal density $\rho(\mathbf{r})$
between the current ($n$) and the previous ($n-1$) iteration steps.
\begin{equation}
	|\Delta F_n|\equiv A^{-1} \int d\mathbf{r}
	\left| \rho^{(n)}(\mathbf{r})-\rho^{(n-1)}(\mathbf{r}) \right| .
\end{equation}
Since the chemical potential is adjusted every iteration
to reproduce either the average nucleon density $\rho_{\rm av}$,
or the nucleon number,
the baryon (nucleon) number $A$ is fixed during the iteration.

In the linear mixing, the result depends on the magnitude
of the mixing parameter $\alpha$.
Figure~\ref{fig:broyden} shows the case of $^{16}$O at $T=1$ MeV.
In this case, the calculation with $\alpha=1$ does not converge,
while that with $\alpha=0.5$ gives the fastest convergence among
$\alpha=1$, 0.8, 0.5, and 0.2.
The optimum value of $\alpha$ varies and is difficult to predict.
For instance, in the case of $T=10$ MeV,
the calculation with $\alpha=1$ converges faster than that with $\alpha=0.5$.
In order to guarantee the convergence,
we need to choose a small value of
$\alpha$, typically $\alpha<0.2$,
which leads to a slow convergence of the iterative procedure.

The modified Broyden's method provides a faster and a stable convergence.
The modified Broyden's algorithm \cite{Joh88} also contains
two parameters we need to choose, namely,
the mixing parameter $\alpha$ and the maximum number of stored vectors $m$.
It turns out that
the result does not strongly depend on the choice of $\alpha$ and $m$.
For the mixing parameter $\alpha$, we can safely choose $\alpha\approx 1$.
Larger values of $m$ give slightly better convergence,
but only a few iteration number difference between $m=10$ and 100.
In the present paper, we adopt $\alpha=0.8$ and $m=100$.
Although storing $m$ Broyden's vectors may require large memory resources when
the system size is large,
the computational time for the Broyden's procedure is negligible.

\subsubsection{Isolated doubly magic nuclei at finite temperature}
\label{sec:closed_nuclei}

\begin{figure}[t]
	\centerline{
	\includegraphics[width=0.5\textwidth]{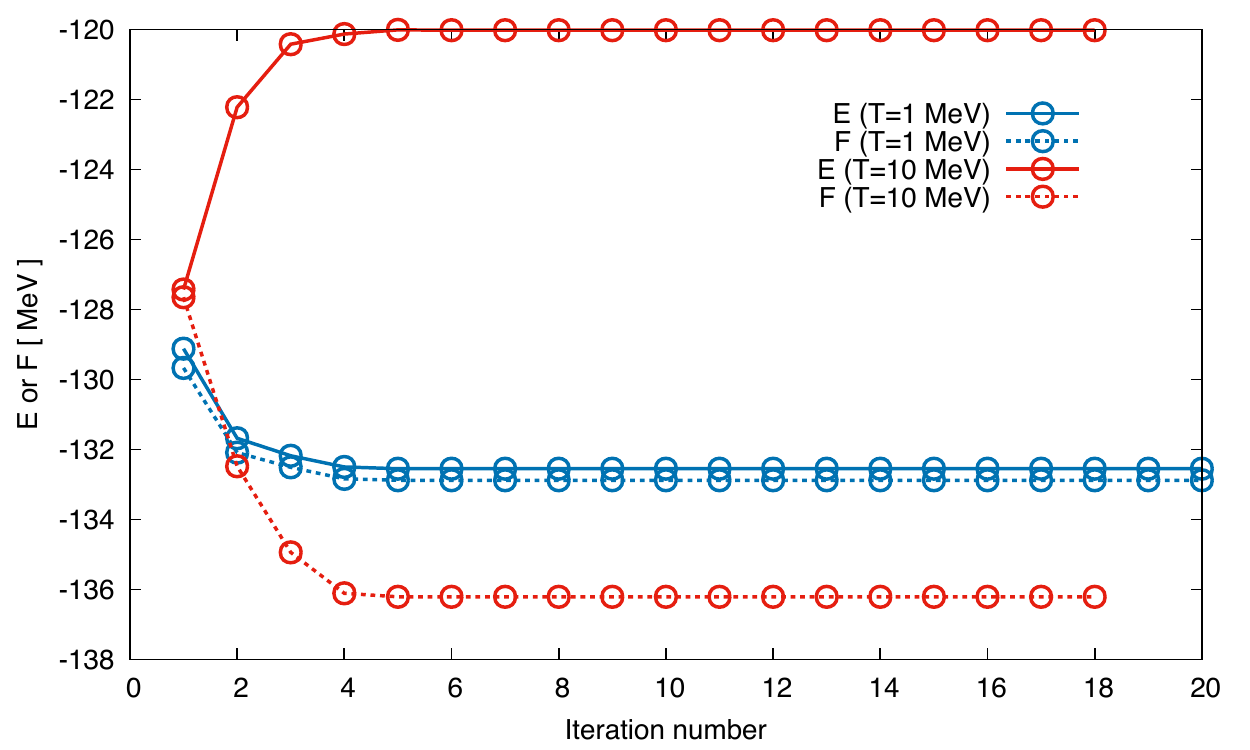}
}
	\caption{
		Total energy $E$ (solid lines) and free energy $F$ (dashed)
		of $^{16}$O at $T=1$ MeV (blue) and $T=10$ MeV (red)
		as functions of self-consistent iteration number
		with the modified Broyden's method.
		The lines for $T=10$ MeV are shifted
		by $-330$ MeV for $E$ and by $+300$ MeV for $F$.
		See text for details.
	}
	\label{fig:16O_E}
\end{figure}

First, let us show results of $^{16}$O at finite temperature.
We use the space size of $(13\mbox{ fm})^3$ with the 3D cubic grid
of $(1\mbox{ fm})^3$.
In Fig.~\ref{fig:16O_E},
the total energy $E$ and the free energy $F$
at every iteration are plotted.
At $T=1$ MeV,
as the iteration number increases,
both $E$ and $F$ decrease to the final values,
$E=-132.6$ MeV and $F=-132.9$ MeV.
The calculated entropy is very small, about 0.3$k_B$.
This is due to the doubly closed-shell nature of $^{16}$O.
At $T=10$ MeV, in contrast,
the total energy $E$ increases to reach the converged value,
$E=210.0$ MeV.
Note that the line in Fig.~\ref{fig:16O_E} is shifted downwards
by 330 MeV to be presented in the same panel as $T=1$ MeV.
Nevertheless, the free energy $F$ decreases,
because the entropy gradually increases as the iteration proceeds.
The entropy is calculated as $S=64.6 k_B$.

\begin{figure}[t]
	\centerline{
	\includegraphics[width=0.5\textwidth]{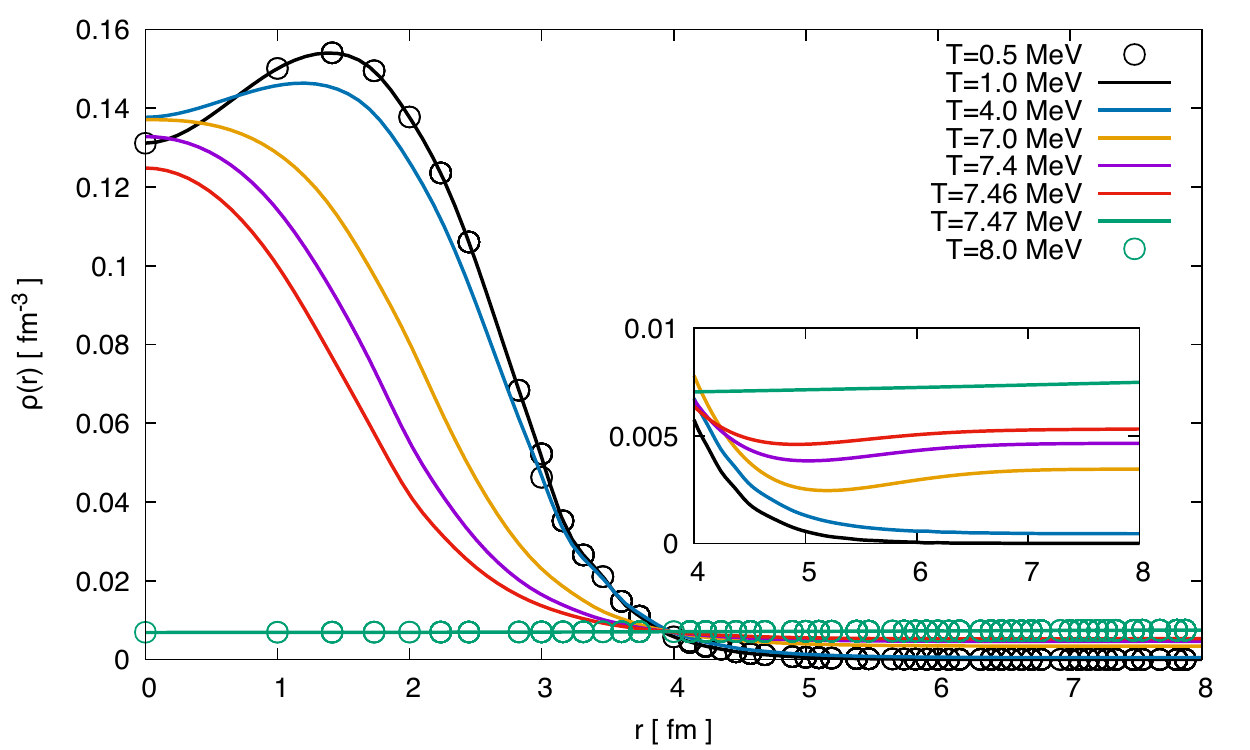}
}
	\caption{
		Nucleon density distribution of $^{16}$O at different temperature.
		The horizontal axis is the distance from the center of mass.
		Since those calculated at $T=0.5$ MeV and 8 MeV
		are indistinguishable
		from those at $T=1$ MeV and 7.47 MeV, respectively,
		they are shown as symbols.
		The space size is $(13\mbox{ fm})^3$.\newline
		Inset: Density distributions
		in the outer region of $r>4$ fm is shown, but
		those at $T=0.5$ MeV and 8 MeV are omitted.
		See text for details.
	}
	\label{fig:16O_rho}
\end{figure}

The calculated nucleon density distributions are presented
in Fig.~\ref{fig:16O_rho}.
The center of mass of $^{16}$O is located
at the center of the cubic box of $(13\mbox{ fm})^3$.
The density values at the calculated grid points are shown
by circles for $T=0.5$ and 8 MeV.
The spline interpolation is used to show the smooth lines
in Fig.~\ref{fig:16O_rho}.
At low temperature, such as $T=0.5$ and 1 MeV,
we find a signature of the shell effect as a dip at the center
of the nucleus.
This is due to the full occupation of the $0p$ orbitals.
Higher is the temperature, more fractional is the occupation,
leading to weakening of the shell effect.
At $T=7$ MeV, the density hole at the center disappears.

The phase transition to the uniformed nuclear matter takes place
at the critical temperature $T_c=7.47$ MeV.
More precisely speaking, it is $7.46<T_c\leq 7.47$ MeV.
A discontinuous change in the density profile
suddenly occurs at $T=T_c$.
This is a consequence of the self-consistent evolution
of the mean-field potential,
which gives a striking contrast to
the density change in the fixed potential (Fig.~\ref{fig:rho_WS}).
The $^{16}$O nucleus is in a liquid phase at $T=0$.
Since there are some dripped nucleons at $T\neq 0$,
it is a coexistence phase of liquid and vapor at $0<T<T_c$.
Then, it is transformed to the gas phase at $T>T_c$.
We should note here that
the critical temperature $T_c$ depends on the adopted volume $V$
that is $(13\mbox{ fm})^3$ in the present calculation.
$T_c$ for the isolated nucleus
should be given as the value at $V\rightarrow\infty$.
See Sec.~\ref{sec:periodic_23fm} for more details.

The dripped nucleons at different temperature can be
seen in the inset of Fig.~\ref{fig:16O_rho}.
In the gas phase, the density should be
$\rho_{\rm gas}=16/(13\mbox{ fm})^3=7.28\times 10^{-3}\mbox{ fm}^{-3}$.
The uniform density obtained at $T>T_c$ is very close to $\rho_{\rm gas}$,
however, the calculated density is not perfectly constant.
It has a minimum value at the center and
slightly increases as $r$ increases.
This strange behavior is an artifact due to the finiteness of the box size.
Following the idea of Ref.~\cite{EB79},
the Coulomb potential for the isolated system
is calculated by assuming that
there exist no charge outside of the adopted space
((13 fm)$^3$ in the present case).
Therefore, the charged particles (protons)
tend to move toward the edge of the box,
in order to reduce the Coulomb repulsive energy.
We have also confirmed that the density is perfectly constant
if the Coulomb potential is neglected.

\begin{figure}[t]
	\centerline{
	\includegraphics[width=0.5\textwidth]{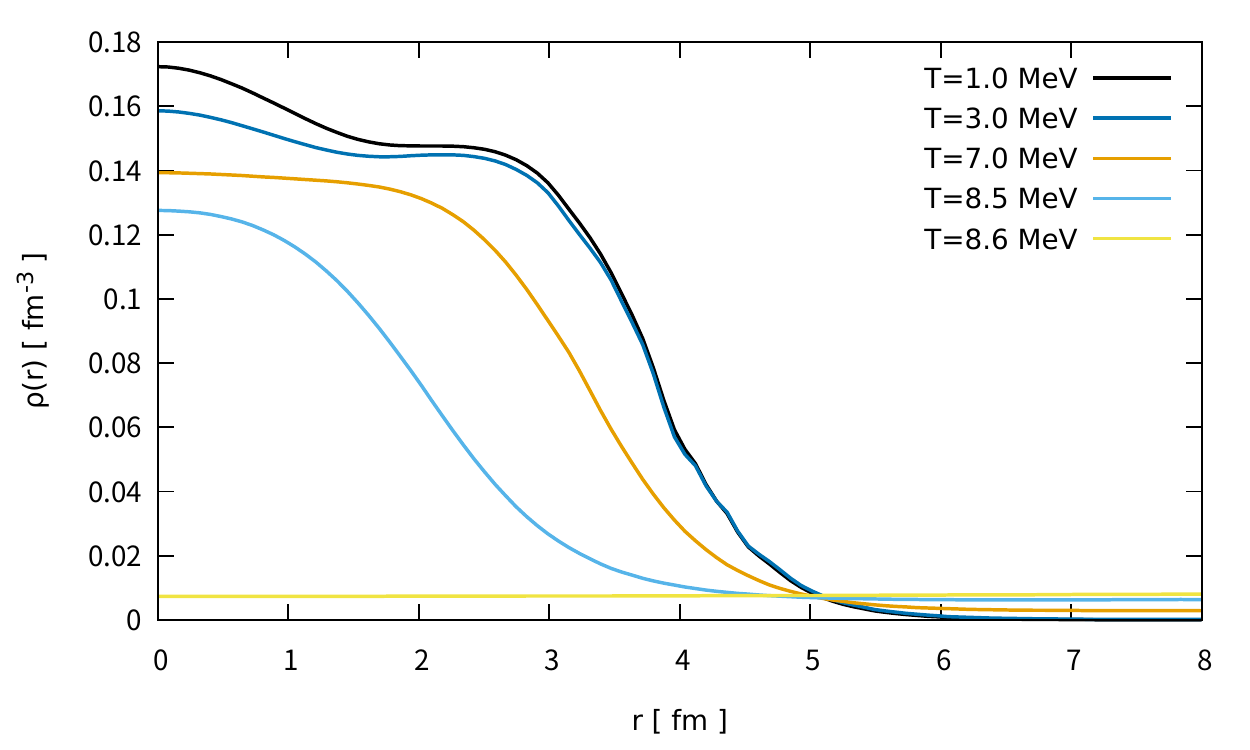}
}
	\caption{
		Nucleon density distribution of $^{16}$O at different temperature.
		The horizontal axis is the distance from the center of mass.
		The space size is $(17\mbox{ fm})^3$.
	}
	\label{fig:40Ca_rho}
\end{figure}

Figure~\ref{fig:40Ca_rho} shows the density profiles for $^{40}$Ca.
The model space is taken as $(17\mbox{ fm})^3$
with the cubic grids of $(1\mbox{ fm})^3$.
The shell effect opposite to $^{16}$O is seen at low temperature,
namely a bump at the center of the nucleus.
This is due to the full occupation of $1s$ orbital.
The shell effect becomes invisible at $T=7$ MeV.
The critical temperature of the liquid-gas transition is located in
$8.5<T_c\leq 8.6$ MeV.
The discontinuous density change is seen at $T=T_c$.
The density of the uniform matter in the present calculation should
be $\rho_{\rm gas}=40/(17\mbox{ fm})^3=8.14\times 10^{-3}$ fm$^{-3}$.

\begin{figure}[t]
	\centerline{
	\includegraphics[width=0.5\textwidth]{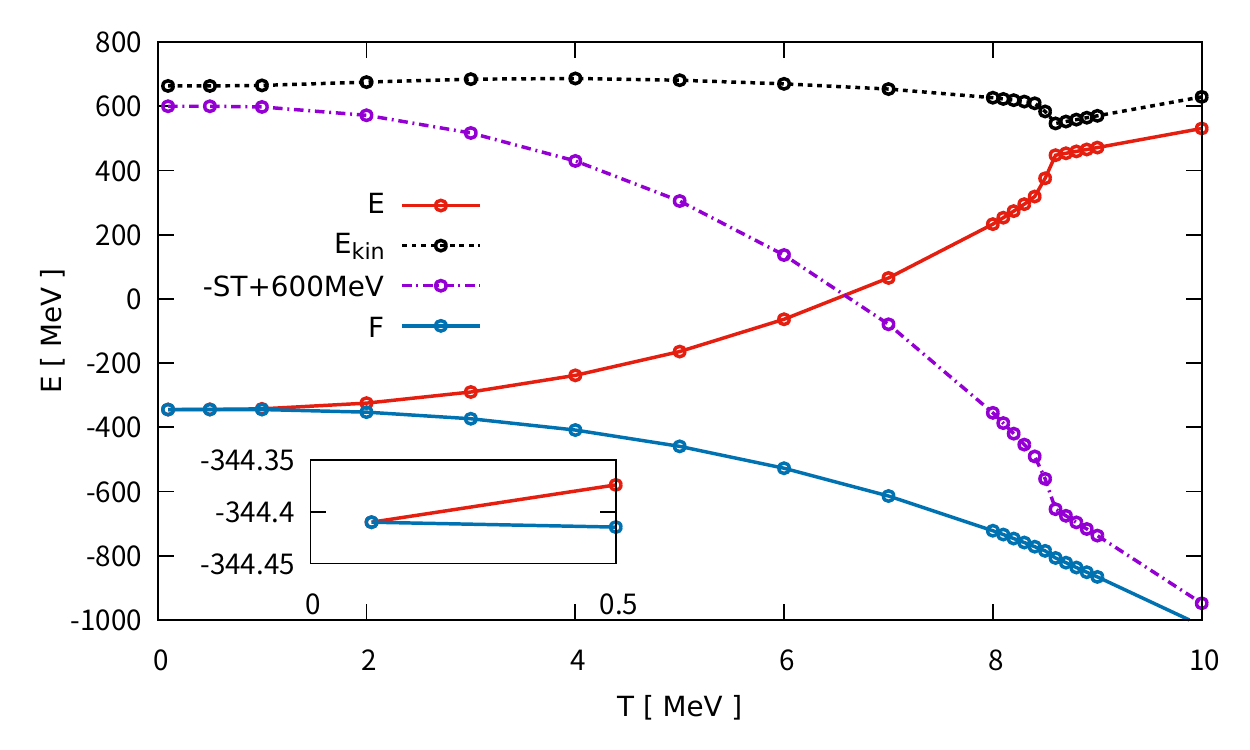}
}
	\caption{
		Total energy $E$ (red solid) and free energy $F$ (blue solid)
		are shown as functions of $T$,
		together with 
		kinetic energy $E_{\rm kin}$ (black dotted) and $-ST$
		(purple dash-dotted)
		for $^{40}$Ca.
		Note that the line of $-ST$ is shifted upward by 600 MeV.\newline
		Inset: $E$ and $F$ in the zero temperature region are magnified.
	}
	\label{fig:40Ca_T-E}
\end{figure}

In Fig.~\ref{fig:40Ca_T-E},
the energy $E$ and the free energy $F$ are shown as functions of $T$.
At $T=T_c\approx 8.6$ MeV, the total energy $E$ shows a kink
because of the abrupt density change.
However, this kink is almost canceled by an opposite kink behavior
in the entropy, and $F=E-ST$ behaves rather smoothly.
$E$ is a monotonic increasing function of $T$, while $F$ is
a decreasing function.
The zero temperature limit is easily achieved in this case,
because the $^{40}$Ca nucleus is doubly magic with large shell gaps
at the Fermi surface.
In the inset panel of Fig.~\ref{fig:40Ca_T-E},
we find that $E=F$ holds in very high accuracy at $T=0.1$ MeV.
The difference is within 0.1 eV.
Even at $T=0.5$ MeV, it is within 50 keV.
Note that the upper (lower) limit of the total energy at $T=0$
is given by $E$ ($F$) at $T>0$.

It is worth mentioning that the coordinate-space representation is
essential to describe the dripped nucleons and the liquid-gas phase transition.
Most of the finite-temperature mean-field calculations in the past
have been performed with the harmonic oscillator basis \cite{Goo81,ER93,ZN18}.
Those studies are focused on the shape change and the pairing properties
at finite temperature, however, it is difficult to describe
the uniform matter and the dripped nucleons.
In Refs.~\cite{BLV84,BLV85},
adopting the spherical Wigner-Seitz approximation,
the finite-temperature Skyrme Hartree-Fock calculation was performed
in the radial coordinate representation.
Our results on properties of the liquid-gas phase transition
turns out to be substantially different from Refs.~\cite{BLV84,BLV85}.
For instance, they showed that
ignoring the Coulomb potential for $^{208}$Pb leads to
a significant increase in $T_c$ (about 5 MeV)
and a smooth continuous transition from
the liquid to the gas phase.
In our calculation,
properties of the liquid-gas transition is almost invariant,
even if we neglect the Coulomb potential:
We find a slight increase by only a few hundreds of keV
with a discontinuous transitions of the density profile
into the uniform matter.
Although the the results in this paper are on $N=Z$ nuclei only,
it would be important to perform the detailed comparison in future
to identify origins of the discrepancies.

\begin{figure}[t]
	\centerline{
	\includegraphics[width=0.4\textwidth]{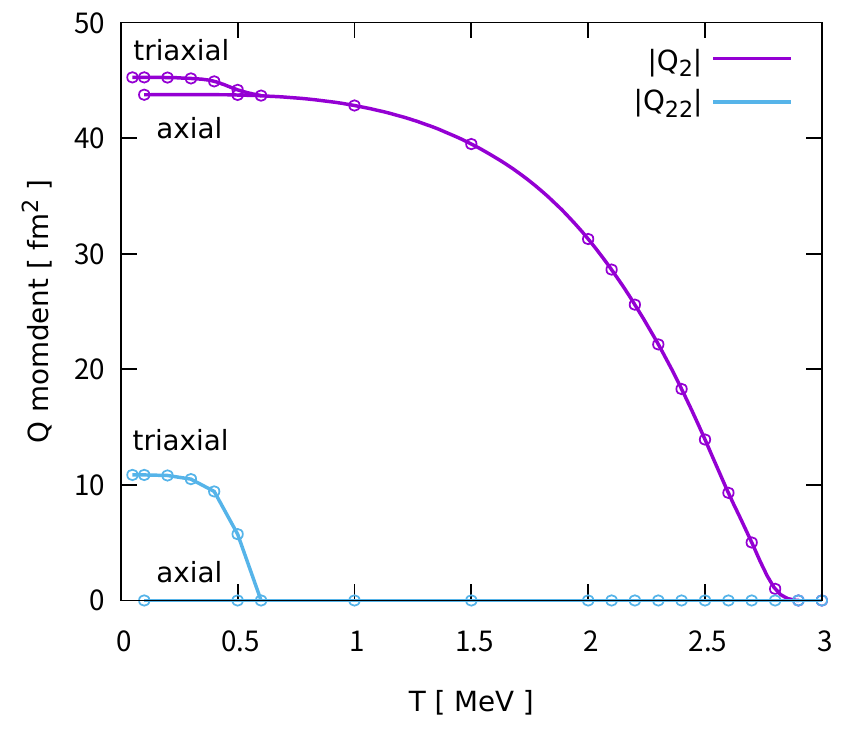}
}
	\caption{
		Calculated intrinsic quadrupole moments 
		($Q_2$ and $|Q_{22}|$)
		for $^{24}$Mg
		at finite temperature.
		At low temperature ($T\lesssim 0.5$ MeV),
		the self-consistent iterations starting from
		different initial states result in different solutions,
		labeled by ``axial'' and ``triaxial'', respectively.
		See text for details.
	}
	\label{fig:24Mg_Q}
\end{figure}

\subsubsection{Isolated deformed nuclei at finite temperature}

We calculate an isolated $^{24}$Mg nucleus
at finite temperature,
which is known to be deformed in the ground state.
In Fig.~\ref{fig:24Mg_Q}, we present the calculated quadrupole moment,
which is defined as
\begin{equation}
	Q_2\equiv \sqrt{\sum_{\mu=-2}^2 |Q_{2\mu}|^2},\quad
	Q_{2\mu}\equiv \int d\mathbf{r} r^2 Y_{2\mu} \rho(\mathbf{r}) .
\end{equation}
At low temperature,
when we start the self-consistent iteration
with a Hamiltonian corresponding to an axially symmetric deformed
density distribution,
the calculation converges to an axially symmetric prolate nucleus
($Q_{2\mu}=0$ except for $\mu=0$).
However, near the zero temperature,
this does not correspond to the state with the minimum free energy.
If we start with a triaxial shape,
it ends up with a triaxially deformed nucleus,
characterized by $Q_{22}\neq 0$.
The shape transition from triaxial to axial shapes takes place
at temperature $T=T_{\rm tri}$ with $0.5 < T_{\rm tri} < 0.6$ MeV.
The axial prolate shape persists till the second shape transition
to the spherical shapes,
which takes place at temperature $T=T_{\rm def}$
with $2.7<T_{\rm def}<2.8$ MeV.
This is shown in Fig.~\ref{fig:24Mg_Q} as two lines.

\begin{figure}[t]
	\centerline{
	\includegraphics[width=0.5\textwidth]{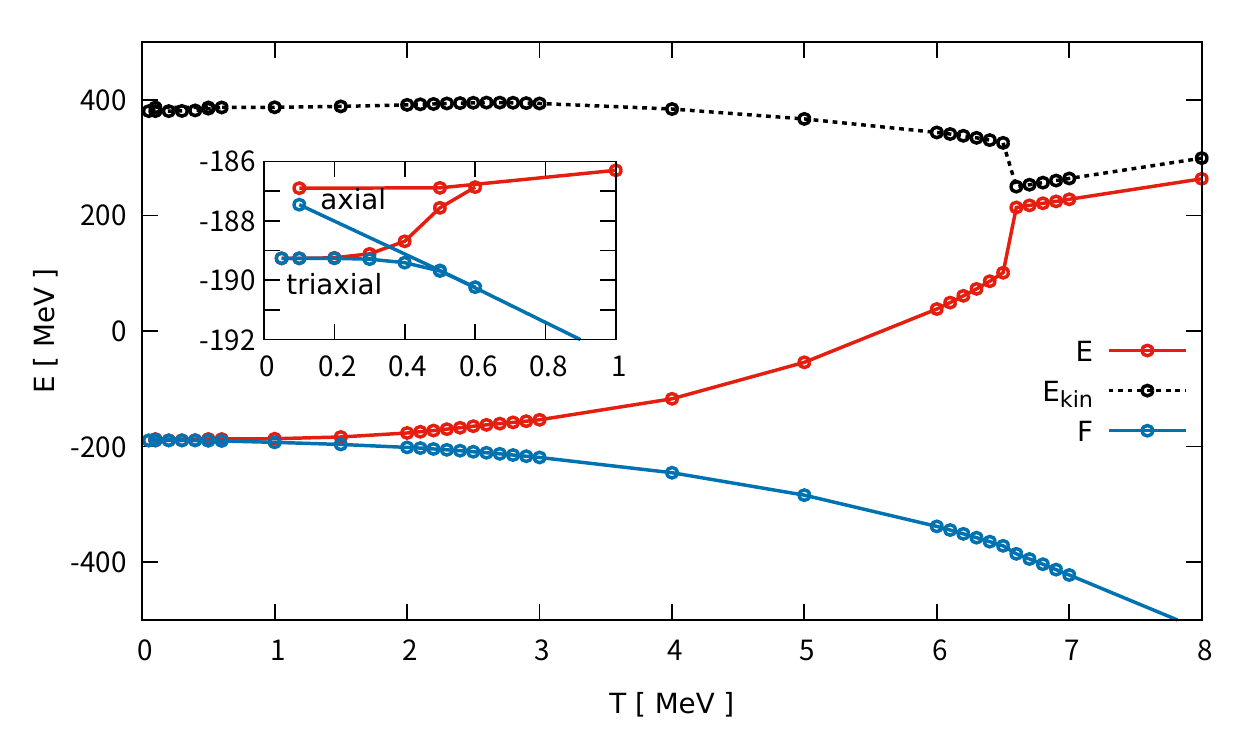}
}
	\caption{
		Total energy $E$ (red solid line)
		and free energy $F$ (blue solid)
		are shown as functions of $T$,
		together with 
		kinetic energy $E_{\rm kin}$ (black dotted)
		for $^{24}$Mg.\newline
		Inset: $E$ and $F$ near the zero temperature are magnified.
	}
	\label{fig:24Mg_E}
\end{figure}

In Fig.\ref{fig:24Mg_E},
we show the temperature dependence of the energy $E$ and the free energy $F$.
A kink of the energy $E$ is caused by the liquid-gas phase transition.
The calculated critical temperature is $6.5<T_c<6.6$ MeV. 
Again, this kink is almost canceled by an opposite kink behavior
in the entropy, and a kink in the free energy $F$ is much smaller.
The effect of the shape transition at $T\approx 2.7$ MeV is
invisible in the temperature dependence of $E$ and $F$,
while that of the axial-triaxial transition at $T\approx 0.5$ MeV
can be seen in the inset of Fig.\ref{fig:24Mg_E}.
An extrapolation to $T=0$ using calculations at $T>0.5$ MeV
may lead to a wrong answer.
The zero temperature limit should be carefully examined
when a structure change is expected at very low temperature.

Another interesting feature is a property of the kinetic energy.
At $T=0.1$ MeV, the kinetic energy for the triaxial solution is
smaller than that of the axial one by about 6.3 MeV,
while the difference in the total energy is about 2.4 MeV.
This clearly indicates that
the triaxial shape in $^{24}$Mg is realized by significant decrease
in the kinetic energy, although it is unfavored by the potential energy.
As the deformation decreases with increasing temperature,
the kinetic energy monotonically increases up to $T=T_{\rm def}$.
This also suggests that the deformation reduces the kinetic energy,
while the potential energy favors the sphericity.
This is consistent with the fact that
the Thomas-Fermi method cannot produce the deformation.
When there are more nucleons moving along $z$ direction than
$x$ and $y$ directions,
according to the uncertainty principle,
the kinetic energy can be reduced by
elongating a potential in the $z$ direction.
This effect is completely lost in 
the local density approximation.

At $T>T_{\rm def}$, the nucleus is spherical and
the kinetic energy decreases as increasing $T$.
At $T=T_c$, since the nucleus suddenly breaks up into a gas phase,
the kinetic energy shows a discontinuous drop.
This is because number of dripped nucleons increases as a function of $T$
up to $T=T_c$,
and their momenta are smaller than those of nucleons confined
inside the nucleus,
which is also due to the uncertain principle.

\subsection{Non-uniform periodic nuclear matter at finite temperature}
\label{sec:periodic_nuclei}

Next, we apply the method to the non-uniform symmetric nuclear matter.
The only difference from the calculations in Sec.~\ref{sec:isolated_nuclei}
is the treatment of the Coulomb potential.
For periodic non-uniform nuclear matter,
we assume the uniform distribution of electrons,
to guarantee the charge neutrality.
This results in the vanishing $k=0$ Fourier component of
the Coulomb potential.
In the present calculations,
the electron energy does not affect the structure of nuclear matter,
since we calculate the nuclear matter at given 
baryon density $\rho$ and proton ratio ($Y_p=0.5$).

\subsubsection{$A=32$ in a cell of $(17\mbox{ fm})^3$}
\label{sec:periodic_17fm}

\begin{figure}[t]
	\centerline{
	\includegraphics[width=0.5\textwidth]{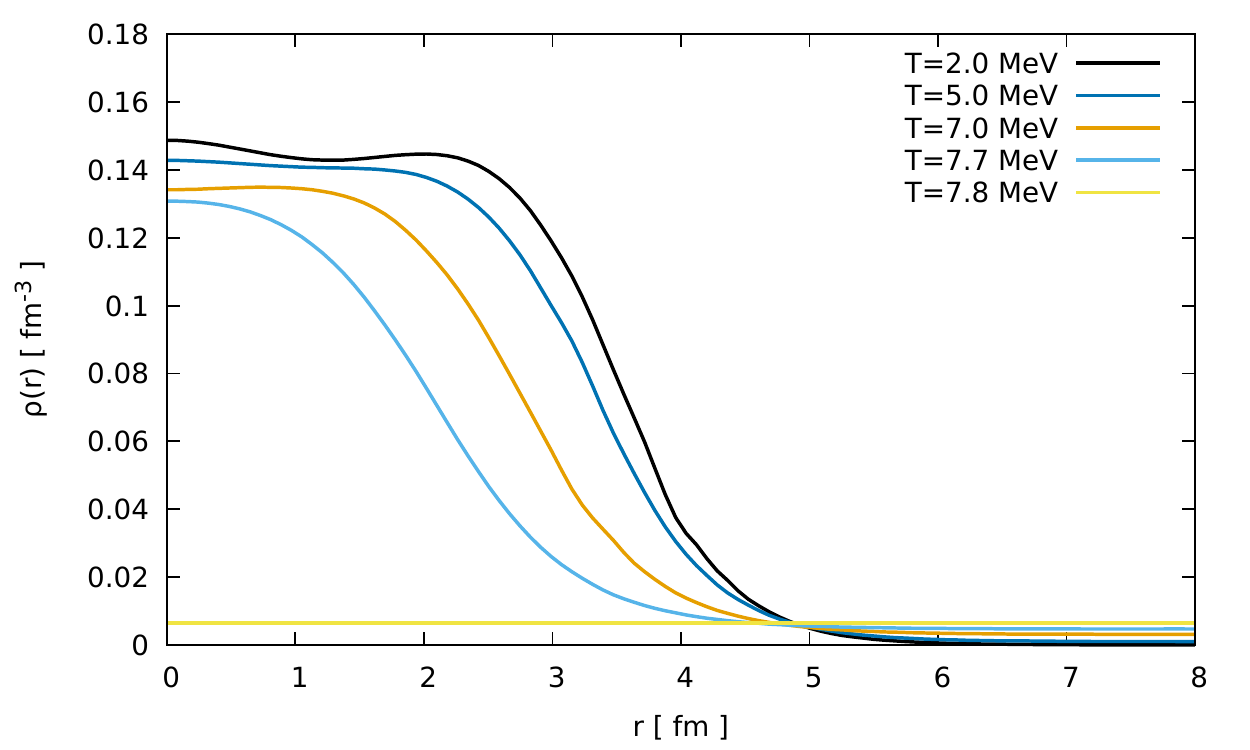}
}
	\caption{
		Density distributions at
                $\rho=6.51\times 10^{-3}$ fm$^{-3}$
		in a cubic configuration at finite temperature.
		The cell size is $(17\mbox{ fm})^3$,
		and the horizontal axis represents the
		distance from the center of the cell.
	}
	\label{fig:32S_cubic}
\end{figure}

First, we calculate the symmetric nuclear matter at average baryon density
$\rho=6.51\times 10^{-3}$ fm$^{-3}$
with a simple cubic initial configuration
in which a $^{32}$S nucleus is located at the center of
a cubic box of $(17\mbox{ fm})^3$.
At low temperature, we find the $^{32}$S nucleus
in a self-consistent solution.
The $^{32}$S nucleus is deformed
at low temperature $T<T_{\rm def}$.
The deformation disappears at $T=T_{\rm def}$
with $1.6<T_{\rm def}<1.7$ MeV.
The dripped nucleons increase with temperature,
then, 
the liquid-gas phase transition takes place at $T=T_c$
with $7.7<T_c<7.8$ MeV.
See Fig.~\ref{fig:32S_cubic} for evolution of the density distribution
as a function of temperature.

\begin{figure}[t]
	\centerline{
	\includegraphics[width=0.5\textwidth]{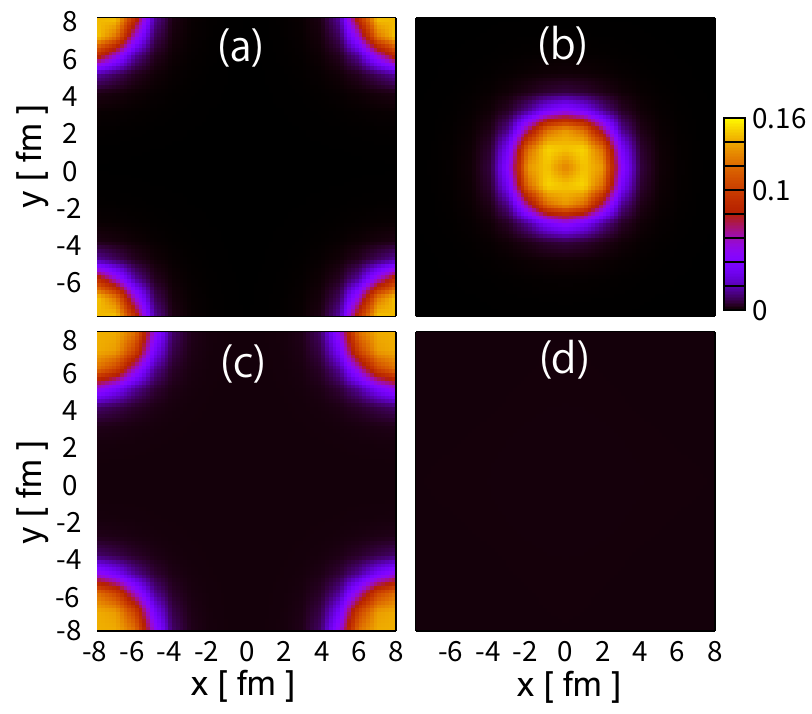}
}
	\caption{
		Density distributions in the $xy$ plane at
                $\rho=6.51\times 10^{-3}$ fm$^{-3}$
		with a cell of $(17\mbox{ fm})^3$,
		calculated with the bcc initial state.
		(a) $T=0.1$ MeV and $z=8$ fm,
		(b) $T=0.1$ MeV and $z=0$,
		(c) $T=5$ MeV and $z=8$ fm,
		(d) $T=5$ MeV and $z=0$.
	}
	\label{fig:32S_bcc}
\end{figure}

In addition to the simple cubic configuration,
we also perform calculations with
the body-centered-cubic (bcc) configuration
as the initial state.
This leads to two $^{16}$O nuclei in the same cell $(17\mbox{ fm})^3$.
At low temperature, the self-consistent calculation converges to
the bcc phase.
Since the (free) energy is larger than that of the cubic configuration,
the bcc state exists as a metastable equilibrium.
Panels (a) and (b) in Fig.~\ref{fig:32S_bcc} show the density distributions
at $T=0.1$ MeV
in the $xy$ plane at $z=8$ fm and at $z=0$.
In contrast, panels (c) and (d) in Fig.~\ref{fig:32S_bcc}
show those at $T=5$ MeV,
indicating that the bcc state is no longer stable
at higher temperature.
During the self-consistent iterations starting from the bcc state,
the $^{16}$O nucleus at the center disappears
leading to the cubic configuration,
namely, a single $^{32}$S nucleus in the cell of
$(17\mbox{ fm})^3$.
The stability of the bcc state seems to be lost around $T=4$ MeV.

Another calculation with the initial configuration
of a $^{40}$Ca nucleus located at the center of the cell
of $(17\mbox{ fm})^3$
is performed.
This corresponds to the average density of
$\rho=8.14\times 10^{-3}$ fm$^{-3}$.
The variation of the density distribution as a function of temperature
is similar to the one for the isolated $^{40}$Ca nucleus
in Fig.~\ref{fig:40Ca_rho}.
However, the critical temperature for the liquid-gas transition
slightly increases, $8.6<T_c<8.7$ MeV.

In the present calculation, 
the number of particles is irrelevant to the computational cost.
Thus, as far as the cell and the grid sizes are invariant,
the computing time is roughly the same for
any density and particle numbers in the cell.
It should be noted that
there is no spurious effect in dripped nucleons,
namely a rise up of the density near the boundary
observed in cases of isolated nuclei
(Sec.~\ref{sec:closed_nuclei})
, such as Fig.~\ref{fig:16O_rho}.
The Coulomb potential in the periodic systems is influenced by
the periodic presence of other nuclei outside of the cell $(17\mbox{ fm})^3$.
The dripped nucleons produce perfectly flat density distribution
outside of the nucleus.

\subsubsection{$A=32$ in a cell of $(23\mbox{ fm})^3$}
\label{sec:periodic_23fm}

\begin{figure}[t]
	\centerline{
	\includegraphics[width=0.5\textwidth]{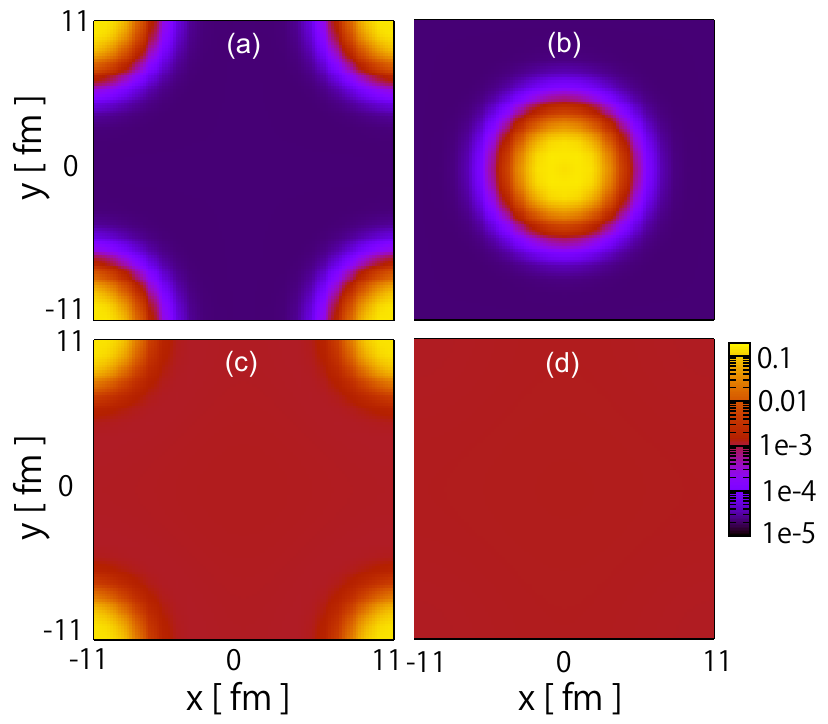}
}
	\caption{
		Density distributions in the $xy$ plane at
                $\rho=2.63\times 10^{-3}$ fm$^{-3}$
		with a cell of $(23\mbox{ fm})^3$,
		calculated with the bcc initial state.
		(a) $T=2$ MeV and $z=11$ fm,
		(b) $T=2$ MeV and $z=0$,
		(c) $T=5$ MeV and $z=11$ fm,
		(d) $T=5$ MeV and $z=0$.
		Note that the color map is given in logarithmic scale.
	}
	\label{fig:32S_23fm}
\end{figure}

Enlarging the cell size into $(23\mbox{ fm})^3$ keeping the baryon number
$A=32$ in the cell,
we perform the same calculations with the cubic and bcc initial configurations.
The average density is $\rho=2.63\times 10^{-3}$ fm$^{-3}$.
Both the cubic and bcc configurations exist at low temperature at $T\lesssim 3.7$ MeV.
The solution with a single $^{32}$S nucleus in the cell 
has lower energy than the bcc solution.
In the calculation with $T\geq 3.8$ MeV, 
the bcc metastable solution seems to disappear, since
we end up with the single $^{32}$S nucleus in the cell
even if we start with the bcc configuration with two $^{16}$O nuclei.
The density profiles obtained with calculations starting from 
the bcc initial configuration are shown in Fig.~\ref{fig:32S_23fm}
at $T=2$ MeV (panels (a) and (b)) and at $T=5$ MeV ((c) and (d)).

\begin{figure}[t]
	\centerline{
	\includegraphics[width=0.5\textwidth]{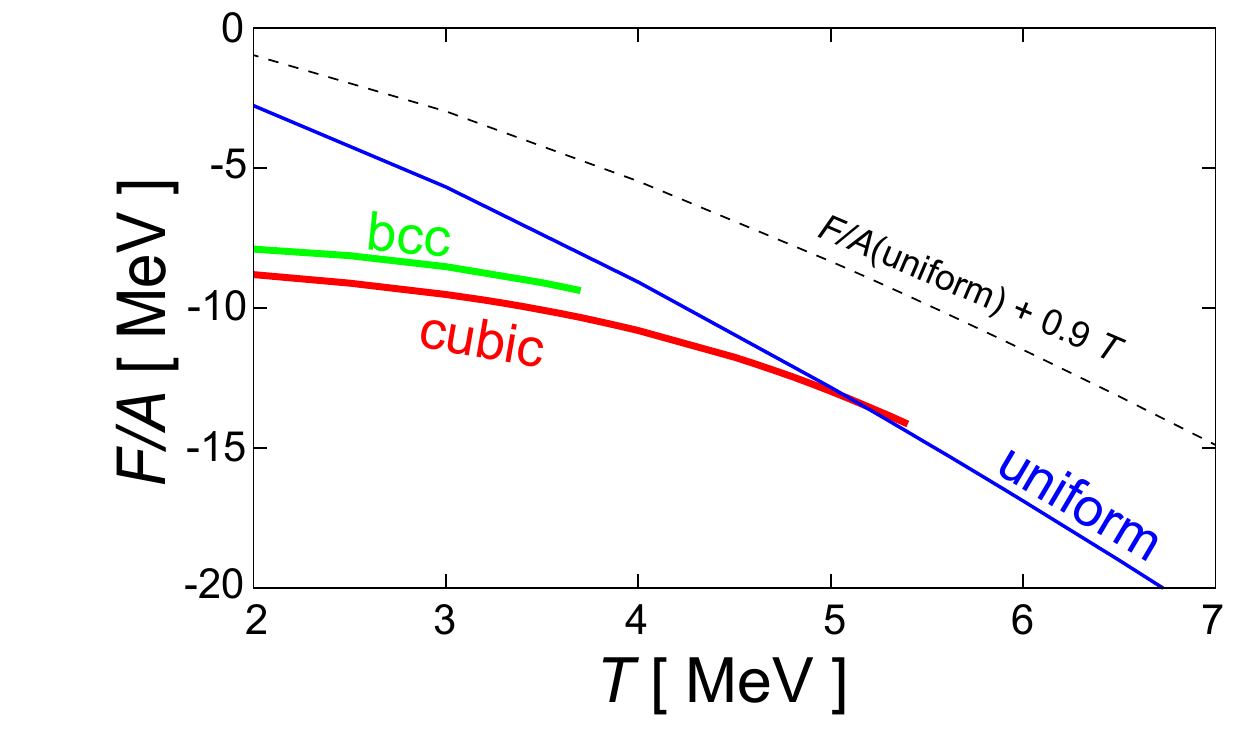}
}
	\caption{
		Free energy per particle of symmetric nuclear matter
		at $\rho=2.63\times 10^{-3}$ fm$^{-3}$
		with a cell of $(23\mbox{ fm})^3$,
		for cubic, bcc, and uniform phases.
		The dashed line is given by shifting
		the line of the uniform matter by $0.9*T$.
		See text for details.
	}
        \label{fig:F/A_32S_23fm}
\end{figure}

Figure~\ref{fig:F/A_32S_23fm} presents the free energy per nucleon
$F/A$ for various phases.
The cubic configuration of $^{32}$S has the lowest free energy
at $T<T_c\approx 5.1$ MeV.
At $T>T_c$, the uniform symmetric matter becomes the lowest.
This critical temperature $T_c$ is significantly smaller than
$T_c$ for the average density
$\rho=6.51\times 10^{-3}$ fm$^{-3}$
with the cell size $(17\mbox{ fm})^3$.
This can be understood as follows:
For the uniform phase at $T>5$ MeV,
the system is well approximated by the classical gas.
The entropy of the classical ideal gas
has the volume dependence as $S\sim Ak_B \ln(V/A)$.
Thus, the entropy per nucleon $S/A$ for the cell of $(23\mbox{ fm})^3$ 
is larger than that of $(17\mbox{ fm})^3$,
by $\delta(S/A)=k_B \ln (23^3/17^3) \approx 0.9 k_B$.
This leads to a shift of $0.9 k_B T$
in the free energy
of the uniform matter in $(17\mbox{ fm})^3$,
shown by a dashed line
in Fig.~\ref{fig:F/A_32S_23fm}.
Since the entropy in the localized phases, such as cubic and bcc,
is scarcely affected by the volume change,
$T_c$, given by the crossing point of the uniform and cubic phases,
decreases as the volume increases.

\subsection{Nearsightedness}
\label{sec:nearsightedness}

Finally, let us check properties of the "nearsightedness" in calculations
of nuclear matter at finite temperature,
then, examine whether it benefits calculations of neutron star matter.
The O($N$) calculation can be achieved if
the one-body density matrix, $\rho(\mathbf{r},\mathbf{r}')$,
is localized in a space considerably smaller than the cell size.
In the present calculation, we can truncate the Hamiltonian matrix
in Eq.~(\ref{eq:recursive_relation_2})
into a space only nearby $\ket{j_n}$.
See also arguments in Sec.~\ref{sec:order-N}.

\begin{figure}[t]
	\centerline{
	\includegraphics[width=0.5\textwidth]{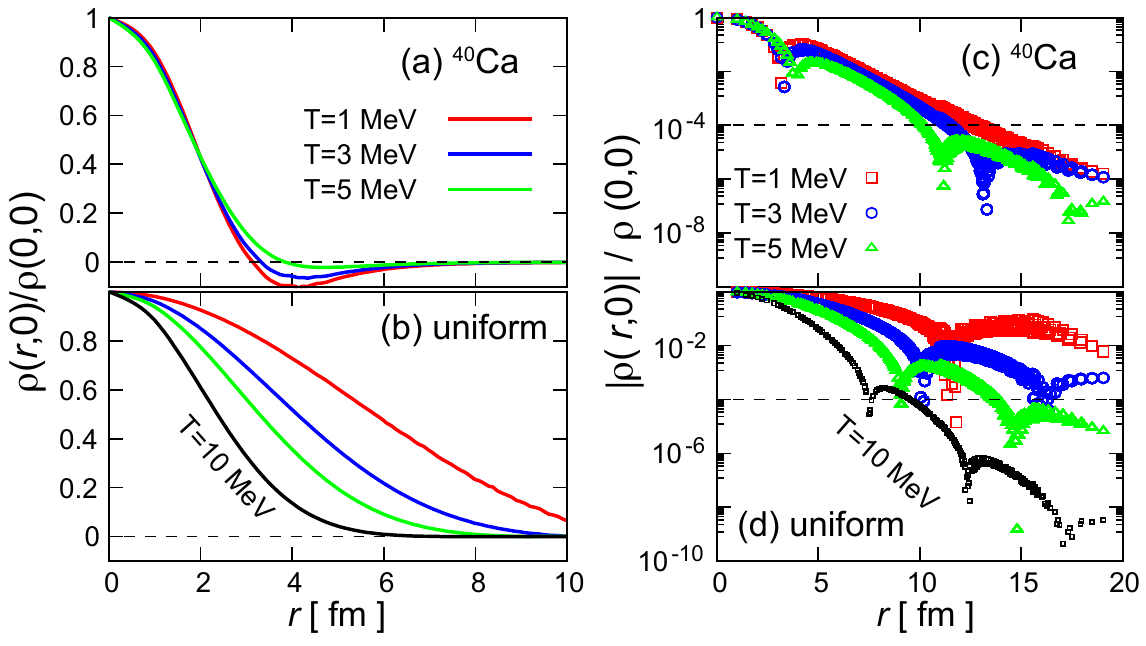}
}
	\caption{
		Normalized density matrix $\rho(r,0)/\rho(0,0)$
		as a function of $r$
		for the simple cubic configuration ((a) and (c))
		and for the uniform matter ((b) and (d)).
		Panels (c) and (d) show the absolute values
		in logarithmic scale,
		with the dashed lines indicating the value of $10^{-4}$.
		The average density is $\rho=8.22\times 10^{-3}$ fm$^{-3}$
		which corresponds to $A=40$ ($^{40}$Ca)
		in a cell of $(23\mbox{ fm})^3$.
	}
        \label{fig:rho_r0}
\end{figure}

Figure~\ref{fig:rho_r0}
presents the off-diagonal behaviors of the density matrix
for a $^{40}$Ca nucleus located at the center of the cell $(23\mbox{ fm})^3$
of the simple cubic lattice.
For comparison, those for the uniform matter is shown in the bottom panels
(b) and (d).
We adopt the center of the cell $\mathbf{r}=0$ as a reference point
and show $\rho(r,0)$ as a function of the distance $r$.
The magnitude of the off-diagonal density exponentially decays.
For the uniform matter,
the calculated behaviors indicate the decay constant proportional to
the temperature $T$.
This is known in studies of finite-temperature density matrix
for electrons in metals \cite{Goe98}.
In contrast, for non-uniform matter with $^{40}$Ca in a cell,
the decay is significantly faster than the uniform matter
and is insensitive to the temperature.
This is not entirely attributed to
the finite radius of the nucleus $^{40}$Ca.
From $T=1$ MeV to $T=5$ MeV,
the radius of $^{40}$Ca is reduced by about $0.5-1$ fm
(cf. Fig.~\ref{fig:40Ca_rho}),
because more nucleons are dripped to form low-density matter.
However, the effect of this reduction in the nuclear radius 
is not visible in Fig.~\ref{fig:rho_r0} (a) and (c).
The fast decay may be, at least partially,
due to large density inside the nucleus.
At zero temperature,
the uniform matter is expected to show
an oscillating pattern of $\rho(r,0)\sim k_F\cos(k_F r)/r^2$ \cite{WJ02},
where $k_F$ is the Fermi momentum.
Thus, at larger density (larger $k_F$),
the off-diagonal density goes to zero more quickly.

Eventually, the localization of the density matrix is 
more prominent in the non-uniform phase than in the uniform matter.
Adopting the cut-off value for the relative magnitude as $10^{-4}$
(the dashed line in Fig.~\ref{fig:rho_r0} (d)),
the cut-off distance for the uniform matter
is given by $R_c\approx 13$ fm at $T=5$ MeV,
and it is considerably larger than 20 fm at $T=1$ MeV.
In contrast, for the non-uniform matter,
the cut-off distance is 
$R_c\approx 10$ fm at $T=5$ MeV
and $R_c\approx 13$ fm at $T=1$ MeV.
When we calculate the $\rho(\mathbf{r},\mathbf{r}')$ with
the recursion relation (\ref{eq:recursive_relation_2}),
truncating the space into a local subspace,
$|\mathbf{r}-\mathbf{r}'|<R_c$,
may lead to a sizable reduction in the computational cost
if the cell size is larger than $R_c^3$.

\section{Conclusion}
\label{sec:conclusion}

We examine the applicability and the usefulness of the Fermi operator expansion
(FOE) method in nuclear energy density functional approaches
at finite temperature.
The one-body density matrix, which is identical to the Fermi operator,
is expanded in terms of the Chebyshev polynomials up to the finite order.
The maximum degree of the polynomials is inversely proportional to the
temperature.
Thus, it becomes extremely efficient for calculations at high temperature.
For the self-consistent iteration procedure,
we adopt the modified Broyden's mixing method.
The same idea of the polynomial expansion is applied
to calculations of the entropy,
which enables us to estimate the free energy without diagonalization of
the Hamiltonian matrix.
The FOE method is applied to
calculations of isolated nuclei and non-uniform nuclear matter,
using the 3D coordinate-space representation.

We investigate thermal properties of isolated nuclei
in a cell of $(13\mbox{ fm})^3$.
For $^{24}$Mg, the triaxial shape has the minimum energy at zero temperature.
The triaxial state exists as a solution at $T\lesssim 0.6$ MeV,
beyond that, the state disappears.
The axial deformed solution survives till $T\approx 2.7$ MeV,
beyond which the nuclear shape is spherical.
The liquid-gas transition takes place around $T_c\approx 6.5$ MeV.
For doubly-magic spherical nuclei, such as $^{16}$O and $^{40}$Ca,
the critical temperature of the liquid-gas transition
has slightly higher values, $T_c=7-9$ MeV.
However, the detailed values of the critical temperature
may not have a significant meaning for isolated nuclei,
because they depends on the volume of the adopted space.
We need to take the infinite volume limit.
Nevertheless, it is a great advantage of
the coordinate-space representation
to be capable of describing both
the spatially localized nucleus and the extended matter.

For periodic non-uniform nuclear matter,
the calculations are performed with different cell sizes,
$(13\mbox{ fm})^3$ and $(23\mbox{ fm})^3$,
with the same nucleon number $A=32$.
We start the self-consistent iteration with
different initial states,
such as the simple cubic and the bcc configurations.
At low temperature, both
the simple cubic and the bcc states exist as
self-consistent solutions.
The cubic state
is lower in free energy than the bcc state.
The transition to the uniform matter takes place at $T_c$,
the value of which is smaller for a larger cell.
This volume effect on the critical temperature $T_c$
is due to the volume dependence of the entropy of the uniform matter.
For the inner crust of neutron stars in the beta equilibrium,
the cell size is supposed to decrease as the density increases
\cite{NV73}.
Since the entropy of the classical gas behaves as
$S/A\sim k_B \ln (V/A) \sim -k_B\ln\rho$,
$T_c$ may become larger at larger densities.
This is somewhat opposite to our naive expectation,
because the density profile becomes flatter at higher density.
It may be of interest to investigate the critical temperature $T_c$
at different density regions.

Advantageous features of the FOE method in computational point of view
can be summarized as follows:
(1) The matrix diagonalization is not involved in the calculation,
including the calculation of the entropy.
(2) The calculation of the density matrix $\rho_{ij}$ is
independent with respect to the index $j$.
Thus, it is suitable for the distributed-memory parallel computing.
(3) The computational cost could scale linearly with respect
to the space dimension $N$,
when $N$ is large enough.
Here, $N$ is the dimension of the matrix $\rho_{ij}$.

The last point (3) above is numerically investigated
by examining the decay of the density matrix $\rho(\mathbf{r},\mathbf{r}')$
at large $|\mathbf{r}-\mathbf{r}'|$.
For the uniform matter, the decay length is shorter at higher temperature,
which has been known for electron systems \cite{WJ02}.
In addition, for non-uniform matter with localized nuclei,
the decay is significantly faster than the uniform matter.
The decay pattern of the non-uniform matter at $T=1$ MeV
is close to that of the uniform matter at $T\approx 10$ MeV.
The short decay length of the density matrix could lead to
the O($N$) calculation by truncating the space
in the matrix operation.
The O($N$) method may be more useful in the non-uniform matter
than in the uniform matter.

The calculations in the present paper use the BKN energy density functional.
It is straightforward to extend this to realistic Skyrme functionals,
which is under progress.
The proper treatment of the matter in the periodic potential
requires the band calculation.
The density should be constructed by averaging the calculated results
over different values of Bloch wave numbers $\mathbf{k}$,
which should be relatively easy to perform.
The calculation can be further parallelized with respect to
different $\mathbf{k}$.

The extension of the FOE method to the finite-temperature HFB calculation
is formally straightforward as well.
This can be done by replacing the single-particle Hamiltonian $H$
by the HFB Hamiltonian,
\begin{equation}
	H_\mu=
	\begin{pmatrix}H-\mu & \Delta \\
	-\Delta^* & -(h-\mu)^* \end{pmatrix} ,
\end{equation} 
in Eq. (\ref{eq:rho}) to achieve the generalized density matrix $R$.
\begin{equation}
	R=\sum_{\alpha\gtrless 0} \ket{\alpha}f_{\beta}(E_\alpha)\bra{\alpha}
	=f_\beta(H_\mu) ,
\end{equation}
where $\ket{\alpha}$ are the quasiparticle eigenstates,
$H_\mu\ket{\alpha}=E_\alpha\ket{\alpha}$,
and the summation is taken over both positive and negative quasiparticle energies.
However, there is a practical issue to be examined in future,
namely,  the truncation of the pairing model space.
Since most of the pairing energy functional has been constructed
with a cut-off energy, it is preferable to develop a prescription to
allow the truncation of the pairing model space.

The FOE method may open a new possibility for studies of
the non-uniform baryonic matter at finite temperature
and neutron-star matter in the crust region.

\begin{acknowledgments}
This work is supported in part by JSPS KAKENHI Grant No. 18H01209.
This research in part used computational resources provided through the HPCI System Research
Project (Project ID: hp200069),
and by Multidisciplinary Cooperative Research Program in Center for
Computational Sciences, University of Tsukuba.
\end{acknowledgments}

\bibliographystyle{apsrev4-1}
\bibliography{nuclear_physics,myself,current}

\end{document}